\documentclass[paper]{JHEP3}
\usepackage{epsfig,subfigure}
\usepackage{amssymb}
\usepackage{amsmath}
\usepackage{booktabs,multirow,tabularx}
\usepackage{slashed}
\renewcommand\arraystretch{1.1}

\newcommand{\bqa}{\begin{eqnarray}}
\newcommand{\eqa}{\end{eqnarray}}

\def\beq{\begin{equation}}
\def\eeq{\end{equation}}
\def\beqn{\begin{eqnarray}}
\def\eeqn{\end{eqnarray}}
\def\abs#1{\left|#1\right|}

\def\MadFKSeq#1{eq.~({\bf I}.#1)}

\newcommand\sss{\scriptscriptstyle}

\newcommand\qb{\bar{q}}
\newcommand\ub{\bar{u}}

\newcommand\epem{e^+e^-}
\newcommand\mpmm{\mu^+\mu^-}

\newcommand\proc{r}

\newcommand\nlight{n_{\sss L}}
\newcommand\nlightB{\nlight^{\sss (B)}}

\newcommand\ident{{\cal I}}

\newcommand\ampsq{{\cal M}}

\newcommand\ampsqnt{\ampsq^{(n,0)}}
\newcommand\ampsqnpot{\ampsq^{(n+1,0)}}

\newcommand\vampsqnlF{{\cal V}^{(n,1)}_{\sss FIN}}

\newcommand\xii{\xi_i}
\newcommand\yij{y_{ij}}

\newcommand\Sfun{{\cal S}}
\newcommand\Sfunij{\Sfun_{ij}}

\newcommand\phsp{d\phi}
\newcommand\phspn{\phsp_{n}}

\newcommand\Phsp{\Phi}
\newcommand\Phspn{\Phsp^{(n)}}
\newcommand\Phspnpo{\Phsp^{(n+1)}}
\newcommand\tPhspnpo{\tilde{\Phsp}^{(n+1)}}
\newcommand\conf{{\cal K}}
\newcommand\confn{\conf_n}
\newcommand\confnpo{\conf_{n+1}}

\newcommand\confnpoE{\conf_{n+1}^{(E)}}
\newcommand\confnpoC{\conf_{n+1}^{(C)}}
\newcommand\confnpoS{\conf_{n+1}^{(S)}}
\newcommand\confnpoSC{\conf_{n+1}^{(SC)}}
\newcommand\confni{\conf_{n;i}}

\newcommand\confnpoai{\conf_{n+1;i}^{(\alpha)}}
\newcommand\confnpoEi{\conf_{n+1;i}^{(E)}}
\newcommand\confnpoSi{\conf_{n+1;i}^{(S)}}

\newcommand\confHi{\conf_{{\sss\clH};i}}
\newcommand\confSi{\conf_{{\sss\clS};i}}

\newcommand\wi{\Xi_i}
\newcommand\wia{\Xi_i^{(\alpha)}}
\newcommand\wiH{\Xi_i^{(\clH)}}
\newcommand\wiS{\Xi_i^{(\clS)}}

\newcommand\as{\alpha_{\sss S}}
\newcommand\asotwopi{\frac{\as}{2\pi}}
\newcommand\gs{g_{\sss S}}

\newcommand\xicut{\xi_{cut}}

\newcommand\deltaI{\delta_{\sss I}}

\newcommand\eikint{{\cal E}}

\newcommand\APdamp{\overline{P}}

\newcommand\JetsB{J^{\nlightB}}

\newcommand\avg{{\cal N}}
\newcommand\symm{\varsigma}

\newcommand\symmnpoij{\symm_{ij}^{(n+1)}}

\newcommand\MadFKS{{\sc\small MadFKS}}

\newcommand\MadLoop{{\sc\small MadLoop}}

\newcommand\muF{\mu_{\sss F}}
\newcommand\muR{\mu_{\sss R}}
\newcommand\muFa{\mu_{\sss F}^{(\alpha)}}
\newcommand\muRa{\mu_{\sss R}^{(\alpha)}}

\newcommand\muFaoQt{\left(\!\frac{\muFa}{\QESa}\!\right)^2}
\newcommand\muRaoQt{\left(\!\frac{\muRa}{\QESa}\!\right)^2}

\newcommand\muRC{\mu_{\sss R}^{(C)}}

\newcommand\muRS{\mu_{\sss R}^{(S)}}

\newcommand\muRSC{\mu_{\sss R}^{(SC)}}

\newcommand\muFp{\mu_{\sss F}^\prime}
\newcommand\muRp{\mu_{\sss R}^\prime}
\newcommand\muFap{\mu_{\sss F}^{\prime{\sss (\alpha)}}}
\newcommand\muRap{\mu_{\sss R}^{\prime{\sss (\alpha)}}}

\newcommand\muFaoQtp{\left(\!\frac{\muFap}{\QESa}\!\right)^2}
\newcommand\muRaoQtp{\left(\!\frac{\muRap}{\QESa}\!\right)^2}
\newcommand\muFEp{\mu_{\sss F}^{\prime(E)}}
\newcommand\muREp{\mu_{\sss R}^{\prime(E)}}

\newcommand\muFSp{\mu_{\sss F}^{\prime(S)}}
\newcommand\muRSp{\mu_{\sss R}^{\prime(S)}}

\newcommand\muFEoQtp{\left(\!\frac{\muFEp}{\QESa}\!\right)^2}
\newcommand\muREoQtp{\left(\!\frac{\muREp}{\QESa}\!\right)^2}
\newcommand\muFSoQtp{\left(\!\frac{\muFSp}{\QESa}\!\right)^2}
\newcommand\muRSoQtp{\left(\!\frac{\muRSp}{\QESa}\!\right)^2}
\newcommand\QESa{Q}

\newcommand\QESCt{Q^2}
\newcommand\QESS{Q}
\newcommand\QESSt{Q^2}

\newcommand\QESSCt{Q^2}
\newcommand\clH{{\mathbb H}}
\newcommand\clS{{\mathbb S}}
\newcommand\fo{f_1}
\newcommand\ft{f_2}
\newcommand\fop{f_1^\prime}
\newcommand\ftp{f_2^\prime}
\newcommand\xo{x_1}
\newcommand\xt{x_2}
\newcommand\xoa{x_1^{(\alpha)}}
\newcommand\xta{x_2^{(\alpha)}}
\newcommand\xoi{x_{1;i}}
\newcommand\xti{x_{2;i}}
\newcommand\xoai{x_{1;i}^{(\alpha)}}
\newcommand\xtai{x_{2;i}^{(\alpha)}}
\newcommand\xoEi{x_{1;i}^{\sss (E)}}
\newcommand\xtEi{x_{2;i}^{\sss (E)}}
\newcommand\xoMCc{x_1^{{\sss (\rm MC},c{\sss )}}}
\newcommand\xtMCc{x_2^{{\sss (\rm MC},c{\sss )}}}
\newcommand\xoMCci{x_{1;i}^{{\sss (\rm MC},c{\sss )}}}
\newcommand\xtMCci{x_{2;i}^{{\sss (\rm MC},c{\sss )}}}
\newcommand\Wa{W^{(\alpha)}}

\newcommand\hWaz{\widehat{W}^{(\alpha)}_0}
\newcommand\hWaF{\widehat{W}^{(\alpha)}_{\sss F}}
\newcommand\hWaR{\widehat{W}^{(\alpha)}_{\sss R}}
\newcommand\hWEz{\widehat{W}^{(E)}_0}
\newcommand\hWEF{\widehat{W}^{(E)}_{\sss F}}
\newcommand\hWER{\widehat{W}^{(E)}_{\sss R}}
\newcommand\hWCz{\widehat{W}^{(C)}_0}
\newcommand\hWCF{\widehat{W}^{(C)}_{\sss F}}
\newcommand\hWCR{\widehat{W}^{(C)}_{\sss R}}
\newcommand\hWSz{\widehat{W}^{(S)}_0}
\newcommand\hWSF{\widehat{W}^{(S)}_{\sss F}}
\newcommand\hWSR{\widehat{W}^{(S)}_{\sss R}}
\newcommand\hWSCz{\widehat{W}^{(SC)}_0}
\newcommand\hWSCF{\widehat{W}^{(SC)}_{\sss F}}
\newcommand\hWSCR{\widehat{W}^{(SC)}_{\sss R}}
\newcommand\hWB{\widehat{W}_{\sss B}}
\newcommand\wMCc{w^{{\sss (\rm MC},c{\sss )}}}

\newcommand\WC{W^{(C)}}

\newcommand\WS{W^{(S)}}

\newcommand\WSC{W^{(SC)}}

\newcommand\sigmaLO{\sigma^{\sss\rm (LO)}}
\newcommand\sigmaNLO{\sigma^{\sss\rm (NLO)}}
\newcommand\sigmaNLOa{\sigma^{{\sss (\rm NLO},\alpha{\sss )}}}
\newcommand\sigmaNLOE{\sigma^{\sss {(\rm NLO},E)}}
\newcommand\sigmaH{\sigma^{\sss (\clH)}}
\newcommand\sigmaS{\sigma^{\sss (\clS)}}

\newcommand\sigmaMCc{\sigma^{{\sss (\rm MC},c{\sss )}}}
\newcommand\MC{{\rm MC}}
\newcommand\resc{{\cal R}}
\newcommand\resca{{\cal R}^{(\alpha)}}
\newcommand\evHi{{\cal E}_{\clH;i}}
\newcommand\evSi{{\cal E}_{\clS;i}}

\newcommand{\PY}{{\sc Pythia}}
\newcommand{\pythiasix}{{\sc Pythia6}}
\newcommand{\herwigpp}{{\sc Herwig++}}

\newcommand{\HW}{{\sc HERWIG}}
\newcommand{\madgraph}{{\sc MadGraph}}
\newcommand{\madloop}{{\sc MadLoop}}
\newcommand{\cuttools}{{\sc CutTools}}
\newcommand{\madfks}{{\sc MadFKS}}
\newcommand{\mcatnlo}{{\sc MC@NLO}}
\newcommand{\amcatnlo}{a{\sc MC@NLO}}
\newcommand{\amcatlo}{a{\sc MC@LO}}
\newcommand{\pt}{p_{\sss T}}
\newcommand{\meas}{\chi}
\newcommand{\kR}{\kappa_{\sss R}}
\newcommand{\kF}{\kappa_{\sss F}}

\preprint{
 CERN-PH-TH/2011-250\\
 CP3-11-32\\
 ZU-TH 19/11\\
 NSF-KITP-11-115\\
 }
\title{Four-lepton production at hadron colliders: aMC@NLO predictions 
       with theoretical uncertainties}

\author{Rikkert Frederix\\
Institut f\"ur Theoretische Physik, Universit\"at Z\"urich,
Winterthurerstrasse 190,\\ CH-8057 Z\"urich, Switzerland
}
\author{Stefano Frixione%
  \thanks{On leave of absence from INFN, Sezione di Genova, Italy.}\\
  PH Department, TH Unit, CERN, CH-1211 Geneva 23, Switzerland\\
  ITPP, EPFL, CH-1015 Lausanne, Switzerland
}
\author{Valentin Hirschi\\
  ITPP, EPFL, CH-1015 Lausanne, Switzerland
}
\author{Fabio Maltoni\\
  Centre for Cosmology, Particle Physics and Phenomenology (CP3)\\
  Universit\'{e} catholique de Louvain,  B-1348 Louvain-la-Neuve, Belgium
}
\author{Roberto Pittau\\
  Departamento de F\'\i sica Te\'orica y del Cosmos y CAFPE, 
  Universidad de Granada
}
\author{Paolo Torrielli\\
  ITPP, EPFL, CH-1015 Lausanne, Switzerland
} 
\abstract{We use aMC@NLO to study the production of four charged leptons at
the LHC, performing parton showers with both \HW\ and \pythiasix.  Our
underlying matrix element calculation features the full next-to-leading order
${\cal O}(\as)$ result and the ${\cal O}(\as^2)$ contribution of the $gg$
channel, and it includes all off-shell, spin-correlation,
virtual-photon-exchange, and interference effects. We present several key
distributions together with the corresponding theoretical uncertainties. These
are obtained through a process-independent technique that allows aMC@NLO
to compute scale and PDF uncertainties in a fully automated way and at no
extra CPU-time cost.} 
\keywords{QCD, NLO Computations, Hadronic Colliders}

\renewcommand\arraystretch{1.1}

\begin{document}

\section{Introduction\label{sec:intro}}
Light charged leptons constitute a particularly clean trigger for high-energy
experiments. It is thus relatively easy to measure, despite their being small,
the cross sections of processes that feature several final-state
leptons. Prominent among such processes are those mediated by a pair of
electroweak vector bosons which, depending on their identities, may give rise
to a missing-energy signature as well. Vector boson pair production is
interesting in at least two respects. Firstly, it is an irreducible background
to Higgs signals, in particular through the $W^+W^-$ and $ZZ$ channels which
are relevant to searches for a standard model Higgs of mass larger than about
140~GeV.  While always smaller than the $W^+W^-$ channel, $ZZ$ decays may 
provide a cleaner signal due to the possibility of fully reconstructing the
decay products of the two $Z$'s. Secondly, di-boson cross sections are quite
sensitive to violations of the gauge structure of the Standard Model, and
hence are good probes of scenarios where new physics is heavy and not
directly accessible at the LHC, yet the couplings in the vector boson 
sector are affected.

The Higgs being such an elusive particle, searches require a combination of
data-driven and theoretical methods in order to overcome irreducible
backgrounds. From the theory viewpoint, this essentially implies
computations as accurate as possible, and ideally able to realistically
reproduce actual experimental events. On the other hand, modifications
to tri-linear gauge couplings will tend to manifest themselves at large
transverse momenta, and therefore this is the region where 
theoretical results have to be reliable. The inclusion of 
next-to-leading order (NLO) QCD corrections into four-lepton
cross section predictions is a minimal, and fairly satisfactory, 
way to fulfill both of the two conditions above.

The aim of this paper is that of studying four-lepton hadroproduction at the
NLO accuracy in QCD, by also adding the (finite) contribution due to $gg$
fusion (which is formally of next-to-next-to-leading order, NNLO, yet
enhanced by the gluon PDFs), and by including the matching with parton showers
according to the MC@NLO formalism~\cite{Frixione:2002ik}; this is done by
means of the framework \amcatnlo\ ~\cite{Frederix:2011zi,Frederix:2011qg} .
Since \amcatnlo\ provides the full automation of 
the matching procedure, and of the underlying matrix element 
computations, we can give results for any four-lepton final
state. In order to be definite, we have chosen to consider
the process:
\beqn
pp\,\longrightarrow\, (Z/\gamma^*)\,(Z/\gamma^*)
\,\longrightarrow\, \ell^+\ell^-\ell^{(\prime)+}\ell^{(\prime)-}\,,
\label{proc}
\eeqn
with $\ell\,,\ell^\prime=e\,,\mu$. This choice is simply motivated
by the fact that $ZZ$ production as currently implemented in the MC@NLO 
package~\cite{Frixione:2010wd} does not feature production spin 
correlation and off-shell effects (which on the other hand are 
included, although in an approximated way, in the case of $W^-W^+$ 
and $W^\pm Z$ production in the codes of ref.~\cite{Frixione:2010wd}), 
and virtual-photon contributions with their interference with the $Z$'s.
The present \amcatnlo\ application remedies to these deficiencies
by computing in an exact manner all relevant matrix elements
(including the contributions of singly-resonant diagrams, which
are potentially relevant to analyses in kinematical regions where
one of the $Z$ bosons is forced to be off-shell), and by including,
as was already mentioned, the $gg$-initiated contribution as well
which, although perturbatively suppressed, can be numerically important
at the LHC owing to its dominant parton luminosity. We remind the
reader that $ZZ$ production at parton level and NLO accuracy in QCD
has been studied for two decades now. The on-shell calculations of
refs.~\cite{Ohnemus:1990za,Mele:1990bq} have been subsequently
improved to include leptonic decays~\cite{Dixon:1998py},
singly-resonant diagrams~\cite{Campbell:1999ah}, and anomalous
couplings~\cite{Dixon:1999di}; these results have been used 
recently to match the NLO computation to parton showers according
to the POWHEG formalism~\cite{Melia:2011tj}. The ${\cal O}(\as^2)$ process
$gg\to ZZ$ was first computed in
refs.~\cite{Dicus:1987dj,Glover:1988fe,Glover:1988rg}. These papers have been
later superseded by the inclusion of off-shell effects and $Z/\gamma^*$
interference~\cite{Binoth:2008pr,Campbell:2011bn}.  As far as \mcatnlo\
results are concerned, and on top of the phenomenological results presented
here, this paper is a first for three reasons:
\begin{itemize}
\item The matching with \pythiasix~\cite{Sjostrand:2006za} for a 
kinematically non-trivial process.
\item The use of \MadLoop~\cite{Hirschi:2011pa} for the computation 
of loop-induced processes (i.e., the contributions of finite 
one-loop amplitudes squared).
\item The computations of scale and PDF uncertainties with 
a reweighting technique.
\end{itemize}
In fact, the use of \PY\ for the shower phase in MC@NLO has been 
limited so far to a proof-of-concept case~\cite{Torrielli:2010aw}, while
the squares of loop amplitudes have been considered in \MadLoop\
only for pointwise tests~\cite{Hirschi:2011pa}, and not in
phenomenological applications. Finally, the reweighting procedure
which we describe in this paper, and that is used to compute the
scale and PDF dependences of the cross section, allows one to
determine the corresponding uncertainties without requiring
any additional computing time. We point out that the capabilities listed in
the three items above are not process dependent and are fully automated. 

This paper is organized as follows:
in sect.~\ref{sec:tecn} we discuss some of the technicalities relevant to 
four-lepton production, and the procedure implemented in \amcatnlo\ 
for the determination of scale and PDF uncertainties.
In sect.~\ref{sec:pheno} we present selected phenomenological
results, and in sect.~\ref{sec:concl} we draw our conclusions.
Further details on scale and PDF dependence computations are
reported in appendix~\ref{sec:wgtNLO}.

\section{Technical aspects of the computation\label{sec:tecn}}
The framework of the \amcatnlo\ programme used for the phenomenology
studies of this paper is unchanged w.r.t.~that employed in
ref.~\cite{Frederix:2011qg}; in particular, the underlying
tree-level computations are performed with \madgraph\ v4~\cite{Alwall:2007st}.
We remind the reader that \amcatnlo\ automates all aspects of an
NLO computation and of its matching with partons showers.
One-loop amplitudes are evaluated with \madloop~\cite{Hirschi:2011pa}, 
whose core is the OPP integrand reduction method~\cite{Ossola:2006us} 
as implemented in \cuttools~\cite{Ossola:2007ax}.  Real contributions 
and the corresponding phase-space subtractions, achieved by means of the FKS 
formalism~\cite{Frixione:1995ms}, as well as their combination with the 
one-loop and Born results and their subsequent integration, are performed by
\madfks~\cite{Frederix:2009yq}, which finally takes care of the MC@NLO 
matching~\cite{Frixione:2002ik} as well.

The novel features of \amcatnlo\ whose results we present here
are the matching with the virtuality-ordered \pythiasix\ shower,
based on ref.~\cite{Torrielli:2010aw}; the possibility
of determining through reweighting the scale and PDF uncertainties
affecting our predictions, which we shall discuss in general
in sect.~\ref{sec:THunc} (with some further technical details
given in appendix~\ref{sec:wgtNLO}); and the use of \MadLoop\
for the computation of the ${\cal O}(\as^2)$ partonic process
\beqn
gg\,\longrightarrow\, (Z/\gamma^*)\,(Z/\gamma^*)
\,\longrightarrow\, \ell^+\ell^-\ell^{(\prime)+}\ell^{(\prime)-}\,.
\label{ggproc}
\eeqn
The matrix elements relevant to this process are UV- and IR-finite, and
therefore the corresponding generated events are treated as an LO sample 
from the viewpoint of matching with parton showers. The amplitudes can 
be straightforwardly computed by \MadLoop. In the current
version, \MadLoop\ assumes that these amplitudes will have to 
be multiplied by Born ones; we have extented the scope of the 
code, in order for it to compute amplitudes squared. 

The fact that the process in eq.~(\ref{ggproc}) is not a genuine
virtual correction, i.e.~it lacks an underlying Born, implies
that event unweighting is essentially a brute-force operation
(as opposed to the normal situation where the presence of an 
underlying Born allows one to pre-determine with good accuracy 
the peak structure of the one-loop matrix elements, thus significantly
increasing the unweighting efficiency). This results in a fairly large
number of calls to the matrix elements of eq.~(\ref{ggproc}) per
unweighted event and, given the computing performances of the 
current version of \MadLoop, renders it very time expensive to obtain
a good-sized unweighted-event sample (say, ${\cal O}(1~{\rm M})$). 
We have therefore opted for an alternative approach. We have generated 
unweighted events using the matrix elements of the Born process
\beqn
u\ub\,\longrightarrow\, (Z/\gamma^*)\,(Z/\gamma^*)
\,\longrightarrow\, \ell^+\ell^-\ell^{(\prime)+}\ell^{(\prime)-}\,,
\label{uufake}
\eeqn
multiplied by the $gg$ parton luminosity (rather than by the $u\ub$
one that would be required if computing the Born contribution proper).
The kinematical configurations so obtained have been used to compute
the matrix elements of eq.~(\ref{ggproc}), which thus provide 
the correct event weights. A similar procedure has been employed in a recent
study on Higgs production via heavy-quark loops and has been shown to work 
well even for more complex final states~\cite{Alwall:2011cy}.

\subsection{Scale and PDF uncertainties\label{sec:THunc}}
Among the parameters which enter a short-distance cross section,
renormalization ($\muR$) and factorization ($\muF$) scales and PDFs play a
special role, since their variations are typically associated with the purely
theoretical uncertainty affecting observable predictions (which, in the case
of PDFs, is not quite correct -- one may say that PDF variations parametrize
the uncertainties not directly arising from the process under study).  It is
therefore fortunate that (the bulk of) the time-consuming matrix element
computations can be rendered independent of scales and PDFs, as opposed to
what happens in the case of other parameters, e.g.~particle masses.  This is
doable thanks to the fact that short-distance cross sections can be written
as linear combinations of scale- and PDF-dependent terms, with coefficients
independent of both scales and PDFs; it is thus possible to compute such
coefficients once and for all, and to combine them at a later stage with
different scales and PDFs at essentially zero cost from the CPU viewpoint.
The crucial (and non-trivial) point is that the noteworthy structure mentioned
before is a feature not only of the parton-level LO and NLO cross sections,
but also of the MC@NLO ones. This implies that from the conceptual point of
view the same procedure for determining scale and PDF uncertaintities can be
adopted in MC@NLO as in LO-based Monte Carlo simulations; for the former we
shall simply need to compute a larger number of coefficients than for the
latter.

In order to illustrate what is done in aMC@NLO, we start from dealing
with LO computations, for which the notation is simpler. Here and in
what follows, the expressions for the short-distance cross sections
are taken from ref.~\cite{Frederix:2009yq}. The fully-differential
cross section is:
\beqn
d\sigmaLO&=&
\fo(\xo,\muF)\ft(\xt,\muF) d\sigma^{(B,n)} J_{Bj} d\meas_{Bj}\,,
\label{fact1}
\\
d\sigma^{(B,n)}&=&\ampsqnt\frac{\JetsB}{\avg}\phspn\,.
\label{sigB}
\eeqn
Here, $d\meas_{Bj}$ denotes the integration measure over the Bjorken $x$'s, 
and $J_{Bj}$ is the (possibly trivial) corresponding jacobian factor.
The quantities on the r.h.s.~of eq.~(\ref{sigB}) are 
the scattering amplitude squared $\ampsqnt$ ({\em including coupling
constants}, and colour/spin average and flux factors), a set of
cuts $\JetsB$ which prevent phase-space singularities from appearing
at this perturbative order, an average factor $\avg$ for identical
final-state particles, and the $n$-body phase space $\phspn$.
We shall write the latter as follows:
\beqn
\phspn=\Phspn\left(\confn(\meas_n)\right) d\meas_n\,,
\label{phsp}
\eeqn
where $d\meas_n$ is a measure that understands the choice of $3n-4$
independent integration variables, and $\confn(\meas_n)$ denotes
the four-momentum configuration associated with a given choice
of the latter. Finally, $\Phspn$ is the phase-space factor 
(including jacobian) that arises from the choice of $d\meas_n$. 
One can then rewrite eq.~(\ref{fact1}) as follows:
\beqn
d\sigmaLO=\fo(\xo,\muF)\ft(\xt,\muF) \gs^{2b}(\muR)
w^{(B,n)} d\meas_{Bj}d\meas_n\,,
\label{fact2}
\eeqn
with
\beqn
w^{(B,n)}=
\frac{\ampsqnt}{\gs^{2b}(\muR)}\frac{\JetsB}{\avg}J_{Bj}\Phspn
\label{wBndef}
\eeqn
and, consistently with ref.~\cite{Frederix:2009yq}, $b$ is the power
of $\as$ implicit in $\ampsqnt$. By construction, the quantity
$w^{(B,n)}$ defined in eq.~(\ref{wBndef}) is independent of
scales and PDFs, and is the first example of the coefficients
mentioned at the beginning of this section. 

When integrating eq.~(\ref{fact1}), one obtains the $N$-event set
\beqn
\Big\{\confni\,,\,\xoi\,,\,\xti\,,\,\wi\Big\}_{i=1}^N\,,
\label{LOset}
\eeqn
with
\beqn
\wi=\frac{d\sigmaLO}{d\meas_{Bj}d\meas_n}\left(\confni,\xoi,\xti\right)\,,
\phantom{aaaaaaaa}
\wi={\rm constant}\,,
\label{wgtLO}
\eeqn
in the case of weighted and unweighted event generation 
respectively\footnote{It should be obvious that the kinematic
configurations and Bjorken $x$'s in eq.~(\ref{LOset}) are not
the same for weighted- and unweighted-event generation. Nevertheless,
we have used an identical notation since no confusion is possible.}.
Given an observable $O$, the cross section in the range
\mbox{$O_{\sss\rm LOW}\le O< O_{\sss\rm UPP}$} (e.g., an
histogram bin) will be
\beqn
\sum_{i=1}^N\Theta\!\left(O\left(\confni\right)-O_{\sss\rm LOW}\right)
\Theta\!\left(O_{\sss\rm UPP}-O\left(\confni\right)\right)
\,\wi
\label{histoLO}
\eeqn
at the hard-subprocess level, and 
\beqn
\sum_{i=1}^N
\Theta\!\left(O\left(\MC(\confni,\xoi,\xti)\right)-O_{\sss\rm LOW}\right)
\Theta\!\left(O_{\sss\rm UPP}-O\left(\MC(\confni,\xoi,\xti)\right)\right)
\,\wi
\label{histoLOMC}
\eeqn
after shower. In eq.~(\ref{histoLOMC}), $\MC(\confni,\xoi,\xti)$ denotes the 
complete final-state configuration obtained by showering the hard subprocess.

When making alternative choices for scales and PDFs, say $\muRp$, $\muFp$,
and $f_i^\prime$, one can simply repeat the procedure outlined above;
this is straightforward, but extremely time consuming if the full scale
and PDF uncertainties have to be determined. Alternatively, one can 
compute eqs.~(\ref{histoLO}) and~(\ref{histoLOMC}) by performing
the replacement
\beqn
\wi\;\;\longrightarrow\;\;\wi\resc_i\,,
\label{wtowR}
\eeqn
where
\beqn
\resc_i=
\fop(\xoi,\muFp)\ftp(\xti,\muFp) \gs^{2b}(\muRp)
w^{(B,n)}(\confni) \Bigg/
\frac{d\sigmaLO}{d\meas_{Bj}d\meas_n}(\confni,\xoi,\xti)\,.
\label{rescLOdef}
\eeqn
The denominator here is computed using eq.~(\ref{fact2}), i.e.~with the
original choices of scales and PDFs. As we shall show in the following,
an analogous equation will hold for both the NLO and MC@NLO cases.
From eq.~(\ref{fact2}) we obtain:
\beqn
\resc_i=
\frac{\fop(\xoi,\muFp)\ftp(\xti,\muFp) \gs^{2b}(\muRp)}
{\fo(\xoi,\muF)\ft(\xti,\muF) \gs^{2b}(\muR)}\,,
\label{rescLO}
\eeqn
which shows explicitly that the computation of $\resc_i$ does not
entail any matrix-element calculation. Therefore, after performing
the bulk of the calculation, i.e.~the determination of the set
in eq.~(\ref{LOset}), one can fill a ``central'' histogram with the
weights defined in eq.~(\ref{wgtLO}), and as many ``variation'' 
histograms as one likes with the weights that appear on the 
r.h.s.~of eq.~(\ref{wtowR}), by simply recomputing the factors
$\resc_i$ for all the different scales and PDFs needed.

The procedure outlined above is exact when applied to
eq.~(\ref{histoLO}), but it implies an approximation in the
case of eq.~(\ref{histoLOMC}). This is because the quantity
\mbox{$\MC(\confni,\xoi,\xti)$} does contain an implicit 
dependence on $f_i$, and these functions are not replaced
by $f_i^\prime$ when using eq.~(\ref{wtowR})
(since such a replacement would imply performing the shower
for each new choice of PDFs). However, this approximation
is usually a very good one (barring perhaps the corners of
the phase space, and in particular the large-rapidity regions).
This is empirically well known, and is due to the fact that the Sudakov
form factors used in the backward evolution of initial-state partons
are sensitive to the ratios of PDFs, whose variation is much smaller
than that of their absolute values (the more so within a PDF-error set,
which is the typical application of the procedure discussed here). 
Furthermore, it should be clear that both scale and PDF uncertainties
are defined using well-motivated but ultimately arbitrary conventions,
and therefore are not quantities that can be computed with arbitrary
precision. 

We now turn to discussing the case of NLO short-distance cross
sections. Equation~(\ref{fact2}) is generalized as follows:
\beqn
d\sigmaNLO&=&\sum_\alpha d\sigmaNLOa\,,
\label{fact3a}
\\
d\sigmaNLOa&=&\fo(\xoa,\muFa)\ft(\xta,\muFa) 
\Wa d\meas_{Bj}d\meas_{n+1}\,,
\label{fact3}
\eeqn
where $\alpha=E$, $S$, $C$, and $SC$ correspond to the contributions
of the fully-resolved configuration (the event), and of its soft, 
collinear, and soft-collinear limits (the counterevents) respectively.
The quantities $\Wa$ will be written as follows:
\beqn
\Wa&=&\gs^{2b+2}(\muRa)\left[
\hWaz+\hWaF\log\!\muFaoQt+\hWaR\log\!\muRaoQt\right]
\nonumber\\*&+&
\gs^{2b}(\muRa)\hWB\delta_{\alpha {\sss S}}\,,
\label{Wadef}
\eeqn
where $\QESa$ is the Ellis-Sexton scale; in eq.~(\ref{Wadef}), this scale
(which is extensively used in the manipulation of the one-loop
contribution -- see ref.~\cite{Frederix:2009yq} for a discussion in
the context of \MadFKS) offers a convenient way to parametrize the 
dependence on the factorization and renormalization scales. As the
notation in eqs.~(\ref{fact3}) and~(\ref{Wadef}) suggests, the values
that such scales assume in the event and counterevents may be different
from each other (however, they must tend to the same value when considering
the relevant infrared limits). Furthermore, eq.~(\ref{fact3}) takes into
account the fact that, when using event projection to write an NLO
partonic cross section (see ref.~\cite{Frixione:2002ik}), the Bjorken
$x$'s of the event and counterevents need not coincide. The last 
term on the r.h.s.~of eq.~(\ref{Wadef}) is the Born contribution,
which in \MadFKS\ can be integrated simultaneously with the other
terms that enter an NLO cross section; having a soft-type kinematics,
it is naturally associated with the soft counterevent. The coefficients
$\hWaz$, $\hWaF$, and $\hWaR$ introduced in eq.~(\ref{Wadef}) are 
the analogues of $w^{(B,n)}$ defined in eq.~(\ref{wBndef}) -- they are
scale- and PDF-independent; their explicit forms are given
in appendix~\ref{sec:wgtNLO}. As far as $\hWB$ is concerned, it is
equal to $w^{(B,n)}$, up to a normalization factor (due to the different
integration measures in eq.~(\ref{fact2}) and~(\ref{fact3})).

The integration of the NLO cross section leads to the set of $N$ weighted
events\footnote{We remind the reader that parton-level, NLO events not
matched with showers cannot be unweighted without the introduction of
an arbitrary and unphysical cutoff.}:
\beqn
%%\Bigg\{\bigcup_{\alpha=E,S,C,SC}
%%\Big\{\confnpoai\,,\,\xoai\,,\,\xtai\,,\,\wia\Big\}
\Bigg\{\Big\{\confnpoai\,,\,\xoai\,,\,\xtai\,,\,\wia\Big\}_{\alpha=E,S,C,SC}
%%\}_{\alpha=E}^{SC}
\Bigg\}_{i=1}^N\,,
\label{NLOset}
\eeqn
with
\beqn
\wia=\frac{d\sigmaNLOa}{d\meas_{Bj}d\meas_{n+1}}
\left(\confnpoai,\xoai,\xtai\right)\,.
\label{wgtNLO}
\eeqn
Equation~(\ref{histoLO}) is generalized as follows:
\beqn
\sum_{i=1}^N\,\sum_\alpha
\Theta\!\left(O\left(\confnpoai\right)-O_{\sss\rm LOW}\right)
\Theta\!\left(O_{\sss\rm UPP}-O\left(\confnpoai\right)\right)
\,\wia\,.
\label{histoNLO}
\eeqn
As is well known, for any given $i$ the weights defined in eq.~(\ref{wgtNLO})
may diverge, but the sum in eq.~(\ref{histoNLO}) is finite for any 
infrared-safe observable. When changing scales and PDFs, we can now
adopt the same procedure introduced before for the LO cross sections,
and define the rescaling factors:
\beqn
&&\resca_i=
\fop(\xoai,\muFap)\ftp(\xtai,\muFap)\Bigg\{
\nonumber\\*&&\phantom{aaa}
\gs^{2b+2}(\muRap)\left[
\hWaz(\confnpoai)+\hWaF(\confnpoai)\log\!\muFaoQtp+
\hWaR(\confnpoai)\log\!\muRaoQtp\right]
\nonumber\\*&&\phantom{aaa\times\Bigg(}
+\gs^{2b}(\muRap)\hWB(\confnpoai)\delta_{\alpha {\sss S}}
\Bigg\}\Bigg/
\frac{d\sigmaNLOa}{d\meas_{Bj}d\meas_{n+1}}(\confnpoai,\xoai,\xtai)\,.
\label{rescNLOdef}
\eeqn
The quantities $\resca_i$ thus defined do not factorize the coefficients 
$\widehat{W}$, and hence have a more complicated form than their LO 
counterpart, eq.~(\ref{rescLO}). Nevertheless, precisely as in the case 
of eq.~(\ref{rescLO}), their computations only require the evaluation
of PDFs and coupling constants for each new choice of PDFs and scales,
the idea being that the coefficients $\widehat{W}$ are computed once and
for all when integrating the partonic cross section, and are stored for
their later re-use in eq.~(\ref{rescNLOdef}). 

We have also understood that the formulae presented above apply to 
the contribution due to one given FKS pair, i.e.~they are proportional 
to a given $\Sfun$ function $\Sfunij$. We remind the reader that while 
the observable cross section is obtained by summing over all $\Sfun$-function
contributions, these are fully independent from each other, and can 
be effectively treated as independent partonic processes.

We finally discuss the case of MC@NLO. The short-distance cross
sections can be written as follows~\cite{Frixione:2002ik}:
\beqn
d\sigmaH&=&d\sigmaNLOE-\sum_c d\sigmaMCc\,,
\label{xsecH}
\\
d\sigmaS&=&\sum_c d\sigmaMCc+\sum_{\alpha=S,C,SC} d\sigmaNLOa\,,
\label{xsecS}
\eeqn
with the Monte Carlo counterterms given by:
\beqn
d\sigmaMCc&=&\fo(\xoMCc,\muF)\ft(\xtMCc,\muF) \gs^{2b+2}(\muR)
\wMCc d\meas_{Bj}d\meas_{n+1}\,.
\label{xsecMC}
\eeqn
The index $c$ in eqs.~(\ref{xsecH}) and~(\ref{xsecS}) runs over all
particles which are colour-connected with the particle that branches.
The latter is identified with the parent particle of the FKS pair
$(i,j)$ which determines the $\Sfun$ function $\Sfunij$ implicit
in $d\sigmaNLOa$. This is sufficient to render eqs.~(\ref{xsecH}) 
and~(\ref{xsecS}) locally finite, as prescribed by the MC@NLO
formalism. Note that the sum over FKS pairs is equivalent to summing over
all possible MC branchings. By integrating eqs.~(\ref{xsecH}) 
and~(\ref{xsecS}) one obtains the sets of events:
\beqn
\Big\{\evHi\,,\,\wiH\Big\}_{i=1}^{N_\clH}\,,
\phantom{aaaaaaaa}
\Big\{\evSi\,,\,\wiS\Big\}_{i=1}^{N_\clS}\,,
\label{HSset}
\eeqn
whose kinematic parts are:
\beqn
\evHi&=&\Big\{\confHi\,,\,\xoEi\,,\,\xtEi\,,\,
\{\xoMCci,\xtMCci\}_c\Big\}\,,
\label{Hset}
\\
%%\evSi&=&\Big\{\confSi\,,\,\{\xoai,\xtai\}_{\alpha=S}^{SC}\,,\,
\evSi&=&\Big\{\confSi\,,\,\{\xoai,\xtai\}_{\alpha=S,C,SC}\,,\,
\{\xoMCci,\xtMCci\}_c\Big\}\,.
\label{Sset}
\eeqn
The final-state kinematic configurations relevant to $\clH$ 
and $\clS$ events are:
\beqn
\confHi=\confnpoEi\,,
\phantom{aaaaaaaa}
\confSi=\confnpoSi\,,
\eeqn
where we have used the fact that in \MadFKS\ and aMC@NLO the phase-space
parametrization is chosen in such a way that the kinematics of the
counterevents coincide (up to the irrelevant unresolved partons;
see ref.~\cite{Frederix:2009yq}). The weights that appear in 
eq.~(\ref{HSset}) are:
\beqn
\wiH=\frac{d\sigmaH}{d\meas_{Bj}d\meas_{n+1}}\left(\evHi\right)\,,
\phantom{aaaaaaaa}
\wiS=\frac{d\sigmaS}{d\meas_{Bj}d\meas_{n+1}}\left(\evSi\right)\,,
\label{wHSdef}
\eeqn
for weighted events, and
\beqn
\abs{\wiH}=\abs{\wiS}={\rm constant}
\eeqn
for unweighted events. One can now define rescaling factors analogous
to those of eqs.~(\ref{rescLOdef}) and~(\ref{rescNLOdef}):
\beqn
\resc_i^{(\clH)}&=&\Bigg\{
\fop(\xoEi,\muFEp)\ftp(\xtEi,\muFEp)\gs^{2b+2}(\muREp)
\nonumber\\*&&\phantom{aaaa}\times\left[
\hWEz(\evHi)+\hWEF(\evHi)\log\!\muFEoQtp+
\hWER(\evHi)\log\!\muREoQtp\right]
\nonumber\\*&&-
\sum_c\fop(\xoMCci,\muFEp)\ftp(\xtMCci,\muFEp) \gs^{2b+2}(\muREp)\wMCc
\Bigg\}\Bigg/
\nonumber\\*&&\phantom{a}
\frac{d\sigmaH}{d\meas_{Bj}d\meas_{n+1}}(\evHi)\,,
\label{rescHdef}
\\
\resc_i^{(\clS)}&=&\Bigg\{
\sum_c\fop(\xoMCci,\muFEp)\ftp(\xtMCci,\muFEp) \gs^{2b+2}(\muREp)\wMCc
\nonumber\\*&&+\!\!\!
\sum_{\alpha=S,C,SC}\fop(\xoai,\muFSp)\ftp(\xtai,\muFSp)\Bigg[
\nonumber\\*&&\phantom{aa}
\gs^{2b+2}(\muRSp)\left(
\hWaz(\evSi)+\hWaF(\evSi)\log\!\muFSoQtp+
\hWaR(\evSi)\log\!\muRSoQtp\right)
\nonumber\\*&&\phantom{aaaaa}
+\gs^{2b}(\muRSp)\hWB(\evSi)\delta_{\alpha {\sss S}}
\Bigg]\Bigg\}\Bigg/
\frac{d\sigmaS}{d\meas_{Bj}d\meas_{n+1}}(\evSi)\,.
\label{rescSdef}
\eeqn
Several observations are in order here. Firstly, the weights defined
in eq.~(\ref{wHSdef}) are finite, at variance with those relevant to the 
computation of the parton-level NLO cross section, eq.~(\ref{wgtNLO}). This 
is due to the properties of the MC counterterms, and to the fact that the
kinematics of the $\clH$ and $\clS$ events is uniquely defined.
Secondly, and as a consequence of the former point, the quantities
relevant to the counterevents are summed together in eq.~(\ref{rescSdef}),
at variance with what happens in eq.~(\ref{rescNLOdef}). Thirdly, in
eqs.~(\ref{rescHdef}) and~(\ref{rescSdef}) the scales entering the 
counterevents have been set equal to the values assumed in the soft
configurations, and those entering the MC counterterms equal to the
corresponding NLO event contributions. Many variants of these choices 
are possible (which is part of the ambiguity in the definition of the
scale dependence in an NLO computation), and those given here are
simply the current defaults in aMC@NLO.

Using the definitions given above, eq.~(\ref{histoLOMC}) is generalized
as follows:
\beqn
&&\sum_{i=1}^{N_\clH}
\Theta\!\left(O\left(\MC(\evHi)\right)-O_{\sss\rm LOW}\right)
\Theta\!\left(O_{\sss\rm UPP}-O\left(\MC(\evHi)\right)\right)
\,\wiH\,\resc_i^{(\clH)}\,,
\label{histoNLOMCH}
\\
&&\sum_{i=1}^{N_\clS}
\Theta\!\left(O\left(\MC(\evSi)\right)-O_{\sss\rm LOW}\right)
\Theta\!\left(O_{\sss\rm UPP}-O\left(\MC(\evSi)\right)\right)
\,\wiS\,\resc_i^{(\clS)}\,.
\label{histoNLOMCS}
\eeqn
Here, the same remark as made in the case of LO-showered predictions
applies, that concerns the PDF dependence implicit in
$\MC(\evHi)$ and $\MC(\evSi)$. There is however a further subtlety
in the case of MC@NLO. When showering the Born contribution to $\clS$ 
events, one cancels the MC counterterm contribution to $\clH$ events. 
In the case of initial-state emissions, we can schematically write 
the contribution to the $\clH$ events obtained by showering the
Born as follows:
\beqn
\frac{\fo(\xo/z)}{\fo(\xo)}\,\Big[\fo(\xo){\rm Born}\Big]\,,
\label{backward}
\eeqn
where we have assumed an emission from leg 1 to be definite, and all
overall factors irrelevant to this discussion have been neglected.
The ratio of PDFs in eq.~(\ref{backward}) is due to the MC branching,
and $[\fo(\xo){\rm Born}]$  is the short-distance contribution.
When $\clS$ events are reweighted with $\resc_i^{(\clS)}$
without changing the PDFs in the shower, eq.~(\ref{backward}) becomes:
\beqn
\frac{\fo(\xo/z)}{\fo(\xo)}\,\Big[\fop(\xo){\rm Born}\Big]\,,
\label{backward2}
\eeqn
and therefore the cancellation of the PDFs with argument $\xo$ does not
take place any longer. Since changing the PDFs in the shower would
amount to replacing $\fo$ with $\fop$ in eq.~(\ref{backward2}),
eq.~(\ref{backward2}) implies a fractional difference w.r.t.~the
use of $\fop$ in both the short-distance and shower computations
equal to:
\beqn
\frac{\fo(\xo/z)}{\fop(\xo/z)}\,
\frac{\fop(\xo)}{\fo(\xo)}\,,
\eeqn
which is a number generally fairly close to one. In fact, what is 
discussed here is precisely the situation one faces when running
MC@NLO and adopting different PDFs in the short-distance calculation 
and shower phases. This is rather commonly done, and no significant
differences are observed w.r.t.~the results obtained with the strictly
correct procedure of using the same PDFs in the two phases.
While this approximation may be undesirable when very accurate results
are mandatory, it is always acceptable when computing the PDF uncertainty,
which is by definition not a precision result. 
Furthermore, it should be kept in mind that the procedure currently used 
for determining PDF ``error'' sets is that of employing parton-level NLO 
cross sections; hence, such PDFs are not guaranteed to serve their purpose 
for those observables for which matched predictions differ significantly 
from the NLO ones (i.e., where resummation effects are important).

\section{Phenomenological results\label{sec:pheno}}
In this section we present some sample results obtained 
with \amcatnlo, its LO counterpart (\amcatlo), and for
parton-level NLO (${\cal O}(\as)$) and LO (${\cal O}(\as^0)$)
cross sections. We also give showered predictions for the $gg$ 
${\cal O}(\as^2)$ contribution. In both \HW~\cite{Corcella:2000bw} 
and \PY, we have switched QED showers off. We consider the production 
of the following leptonic final states:
\beqn
\ell^+\ell^-\ell^{\prime+}\ell^{\prime-}&=&\epem\mpmm\,,
\label{eemumu}
\\
\ell^+\ell^-\ell^+\ell^-&=&\epem\epem\,,
\label{eeee}
\eeqn
at the 7~TeV LHC. The only cut at the generation level is:
\beqn
M(\ell^\pm\ell^{(\prime)\mp})\ge 30~{\rm GeV}\,,
\label{Mcut}
\eeqn
which is applied to {\em all} unlike-sign lepton pairs. This cut,
that serves the purpose of avoiding the zero-virtuality divergences
of diagrams with resonant photons, is obviously not necessary
in the case of eq.~(\ref{eemumu}) for $e^\pm\mu^\mp$ pairs, but it
is nonetheless applied there in order to perform a comparison on equal
footing between the $\epem\mpmm$ and $\epem\epem$ final states.
The parameters used in this paper are reported in table~\ref{tab:params},
where the values of $\as$ have been chosen as prescribed by the
PDF sets we have adopted, MSTW2008(n)lo68cl~\cite{Martin:2009iq}
(with their associated ``error'' sets).
The default renormalization and factorization scales
are set equal to the invariant mass of the four leptons, 
\beqn
\mu_0=M(\ell^+\ell^-\ell^{(\prime)+}\ell^{(\prime)-})\,.
\label{defscale}
\eeqn
When studying scale dependences we shall vary $\muR$ and $\muF$
independently, i.e.~we shall set $(\muR,\muF)=(\kR\mu_0,\kF\mu_0)$
and consider the combinations $(\kR,\kF)=(1,1)$, $(1/2,1/2)$, $(1/2,1)$, 
$(1,1/2)$, $(1,2)$, $(2,1)$, $(2,2)$. The overall scale uncertainty will 
be the envelope of all individual results.
%%%%%%%%%%%%%%%%%%%%%%%%%%%%%%%%%%%%%%%%%%%%%%%%%%%%%%%%%%%%%%%%%%%%%%%%
\begin{table}
\begin{center}
\begin{tabular}{ll|ll}\toprule
Parameter & value & Parameter & value
\\\midrule
$m_{Z}$ & 91.118 & $\Gamma_Z$ & 2.4414 
\\
$\alpha^{-1}$ & 132.50698  & $G_F$ & $1.16639\!\cdot\!10^{-5}$
\\
$\as^{({\rm LO})}(m_{Z})$  & 0.13939 & $\as^{({\rm NLO})}(m_{Z})$  &  0.12018
\\\bottomrule
\end{tabular}
\end{center}
\caption{\label{tab:params}
  Settings of physical parameters used in this work,
  with dimensionful quantities given in GeV.
}
\end{table}
%%%%%%%%%%%%%%%%%%%%%%%%%%%%%%%%%%%%%%%%%%%%%%%%%%%%%%%%%%%%%%%%%%%%%%%%

We begin by presenting in table~\ref{tab:xsec} the fully inclusive
cross sections (bar the invariant-mass cuts of eq.~(\ref{Mcut})).
For the NLO results and the $gg$ contribution we also give the scale
(first error) and PDF (second error) uncertainties. The scale 
dependence at the NLO is fairly small, which is in part due to the 
fact that the $q\qb$ parton luminosity is almost $\muF$-independent 
in the Bjorken-$x$ range that gives the dominant contribution to the 
cross section. This implies that at the LO (which, being ${\cal O}(\as^0)$, 
is independent of $\muR$) the scale uncertainty is comparable or smaller,
and thus not particularly meaningful: we refrain from quoting it explicitly
here. The PDF uncertainty is estimated following the prescription given
by the PDF set~\cite{Martin:2009iq} (asymmetric Hessian); at the LO, we 
obtain a fractional uncertainty similar to that of the NLO result. On 
the other hand, the $gg$ channel displays the typical scale dependence of
an ${\cal O}(\as^2)$ LO cross section, the figures in the table 
receiving the dominant contributions from the $\muR$ dependence.
The scale uncertainty is largely dominant over the PDF one.
Not surprisingly, the ratio between the two leptonic channels considered
in this paper is equal to two up to a very small correction (only slightly
larger than the integration uncertainty), since differences arise 
in the off-resonance regions (where singly-resonant diagrams are
more important than elsewhere) whose contribution to the total
integral is small. Given that the $gg$ matrix elements do feature only
doubly-resonant diagrams, we have not computed the result relevant
to the $\epem\epem$ channel. Finally, we point out that the entries
of table~\ref{tab:xsec} have been compared with their analogues
obtained from MCFM (also for the non-central scale values, which is a 
check of the reweighting technique discussed in sect.~\ref{sec:THunc}), 
and complete agreement has been found.
%%%%%%%%%%%%%%%%%%%%%%%%%%%%%%%%%%%%%%%%%%%%%%%%%%%%%%%%%%%%%%%%%%%%%%%%
\begin{table}[t]
\begin{center}
\renewcommand*{\arraystretch}{1.5}
\begin{tabular}{|c|cc|c|}
\hline
 & \multicolumn{3}{c|}{Cross section (fb)}\\
\cline{2-4}  
 Process  & \multicolumn{2}{c|}{$q\qb$/$qg$ channels} & {$gg$ channel} \\
\cline{2-4}
& ${\cal O}(\as^0)$ & ${\cal O}(\as^0)+{\cal O}(\as)$ & ${\cal O}(\as^2)$ \\
\hline
$pp\to\epem\mpmm$  &   
% 9.193 & $12.90^{+0.27(2.1\%)+0.26(2.0\%)}_{-0.23(1.8\%)-0.22(1.7\%)}$  & 
 9.19 & $12.90^{+0.27(2.1\%)+0.26(2.0\%)}_{-0.23(1.8\%)-0.22(1.7\%)}$  & 
 $0.566^{+0.162(28.6\%)+0.012(2.1\%)}_{-0.118(20.8\%)-0.014(2.5\%)}$     \\
$pp \to e^+ e^- e^+ e^-$      &   
%% 4.577 & $6.435^{+0.133(2.1\%)+0.111(1.7\%)}_{-0.130(2.0\%)-0.100(1.6\%)}$&\\
 4.58 & $6.43^{+0.13(2.1\%)+0.11(1.7\%)}_{-0.13(2.0\%)-0.10(1.6\%)}$ & \\
\hline
\end{tabular}
\end{center}
\caption{Total cross sections for $\epem\mpmm$ and $\epem\epem$
production at the LHC ($\sqrt{S}= 7$~TeV), for the production 
channels considered in this paper, and within the cuts given in
eq.~(\ref{Mcut}). The first and second errors affecting the results
are the scale and PDF uncertainties (also given as fractions of the
central values). See the text for details.}
\label{tab:xsec}
\end{table}
%%%%%%%%%%%%%%%%%%%%%%%%%%%%%%%%%%%%%%%%%%%%%%%%%%%%%%%%%%%%%%%%%%%%%%%%

As was mentioned above, the scale uncertainties reported in
table~\ref{tab:xsec} are the envelopes of the results obtained
with the seven combinations of $(\muR,\muF)$ considered here.
It is obvious that this is not equivalent to variations
in the range $\mu_0/2\le\muR,\muF\le 2\mu_0$ if the dependence
on either scale is not monotonic. While the study of continuous
variations is clearly impossible, by means of reweighting one
can probe the relevant scale ranges with more than three
values for each scale, at a very modest CPU cost. In fact, each
choice of $(\muR,\muF)$ corresponds to filling an extra set of
histograms (each set containing all observables of interest), with 
the same kinematics as for the central set and with rescaled weights, 
which is a negligible fraction of the generation-plus-shower computing time.

\begin{figure}[t]
\centering
\includegraphics[width=0.49\textwidth]{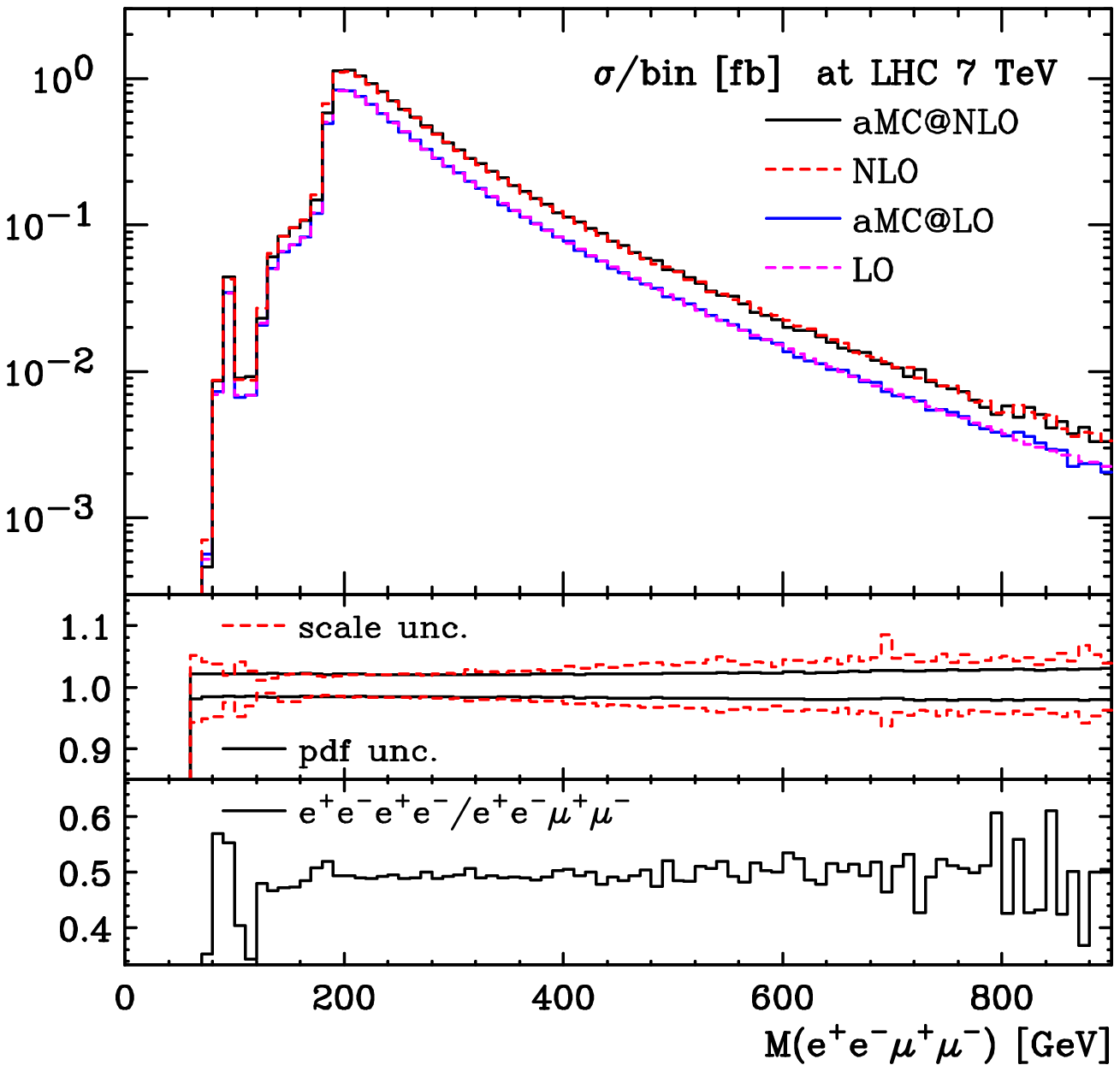}
\includegraphics[width=0.49\textwidth]{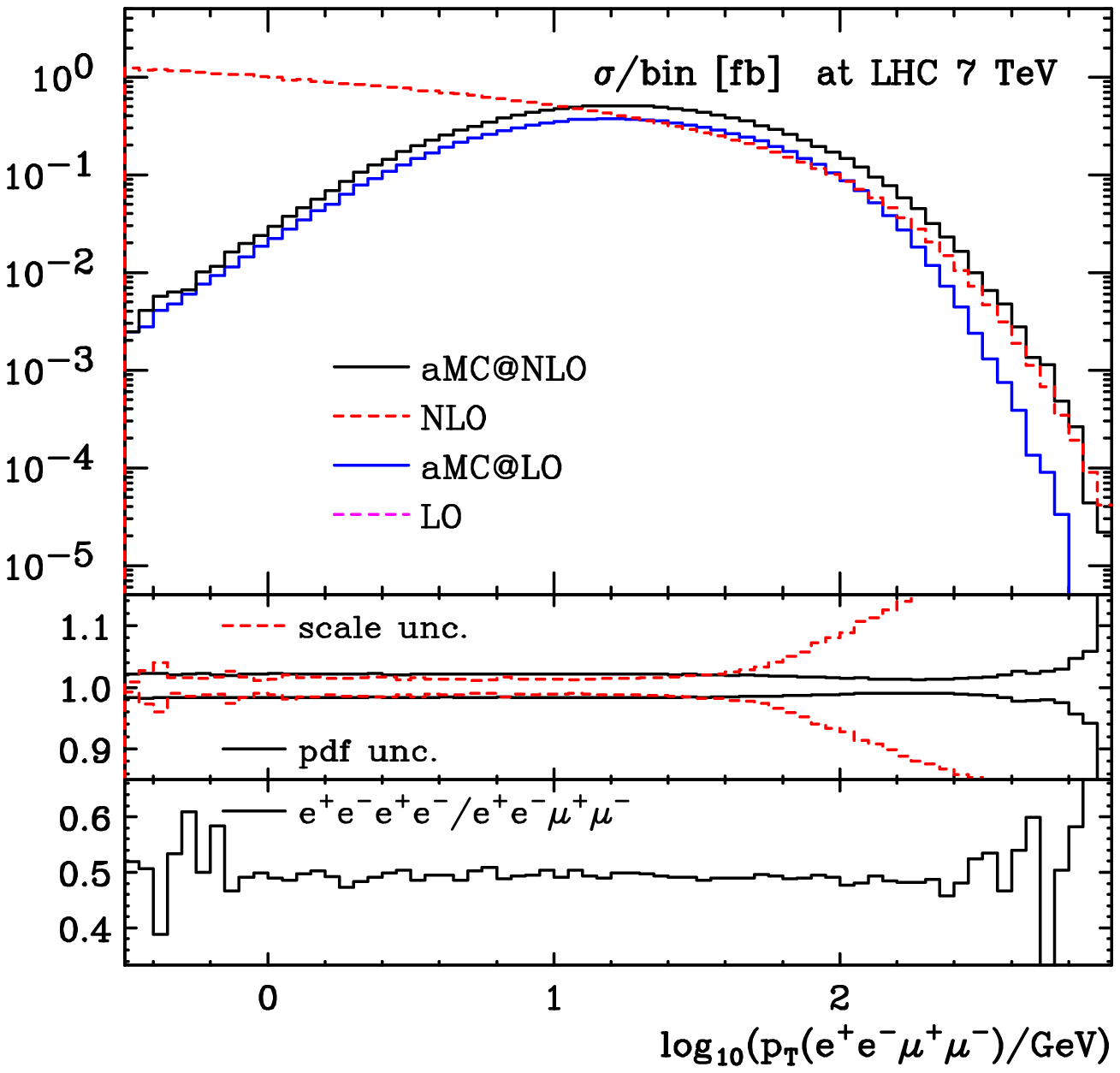}
\caption{Four-lepton invariant mass (left panel) and transverse 
momentum (right panel), as predicted by \amcatnlo (solid black), \amcatlo 
(solid blue), and at the (parton-level) NLO (dashed red) and LO (dashed 
magenta).  The middle insets show the \amcatnlo\ scale (dashed red) and PDF 
(black solid) fractional uncertainties, and the lower 
insets the ratio of the two leptonic channels, eq.~(\ref{ratofchs}).
See the text for details.}
\label{fig4:rad}
\end{figure}

\begin{figure}[t]
\centering
\includegraphics[width=0.49\textwidth]{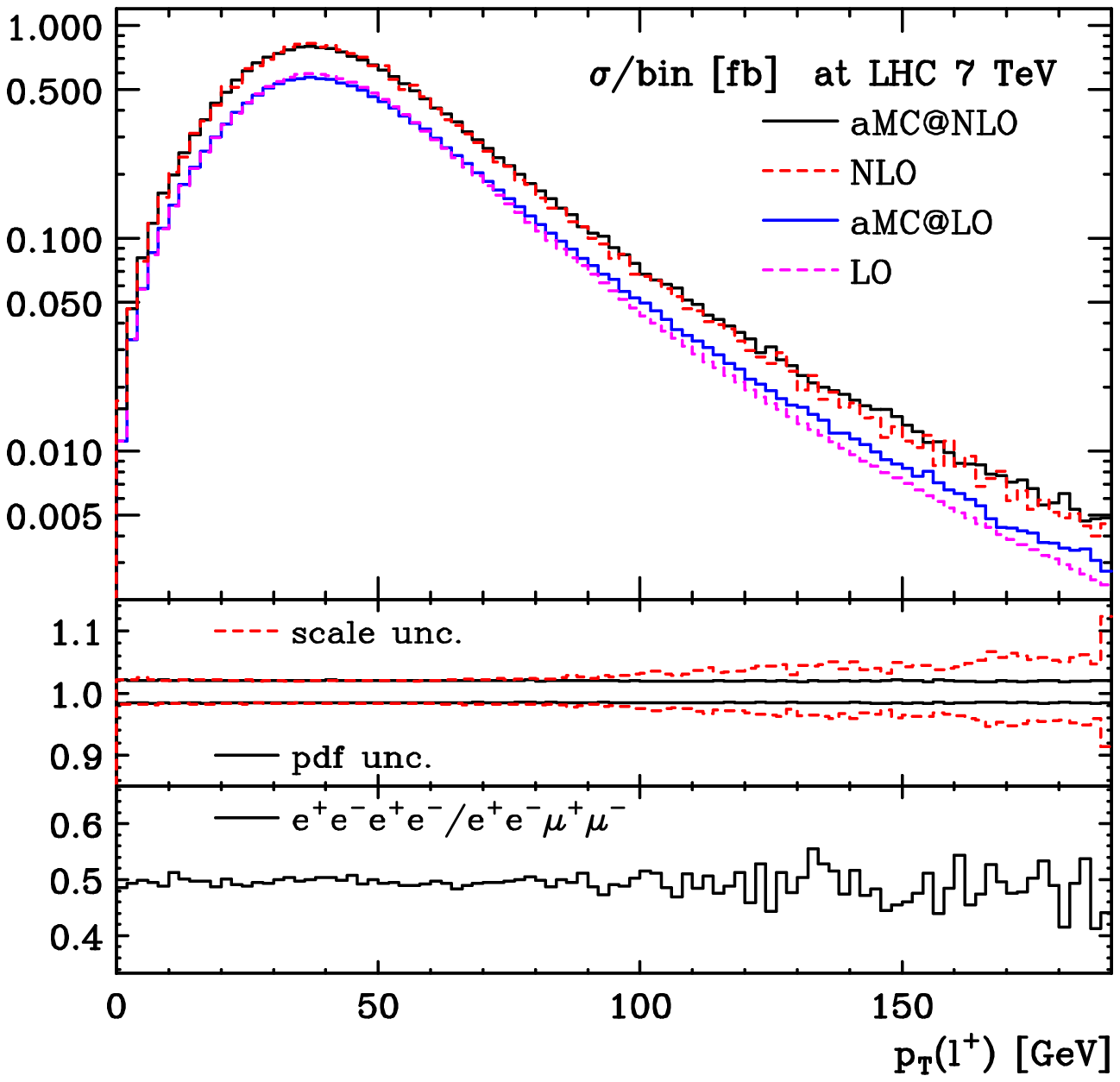}
\includegraphics[width=0.49\textwidth]{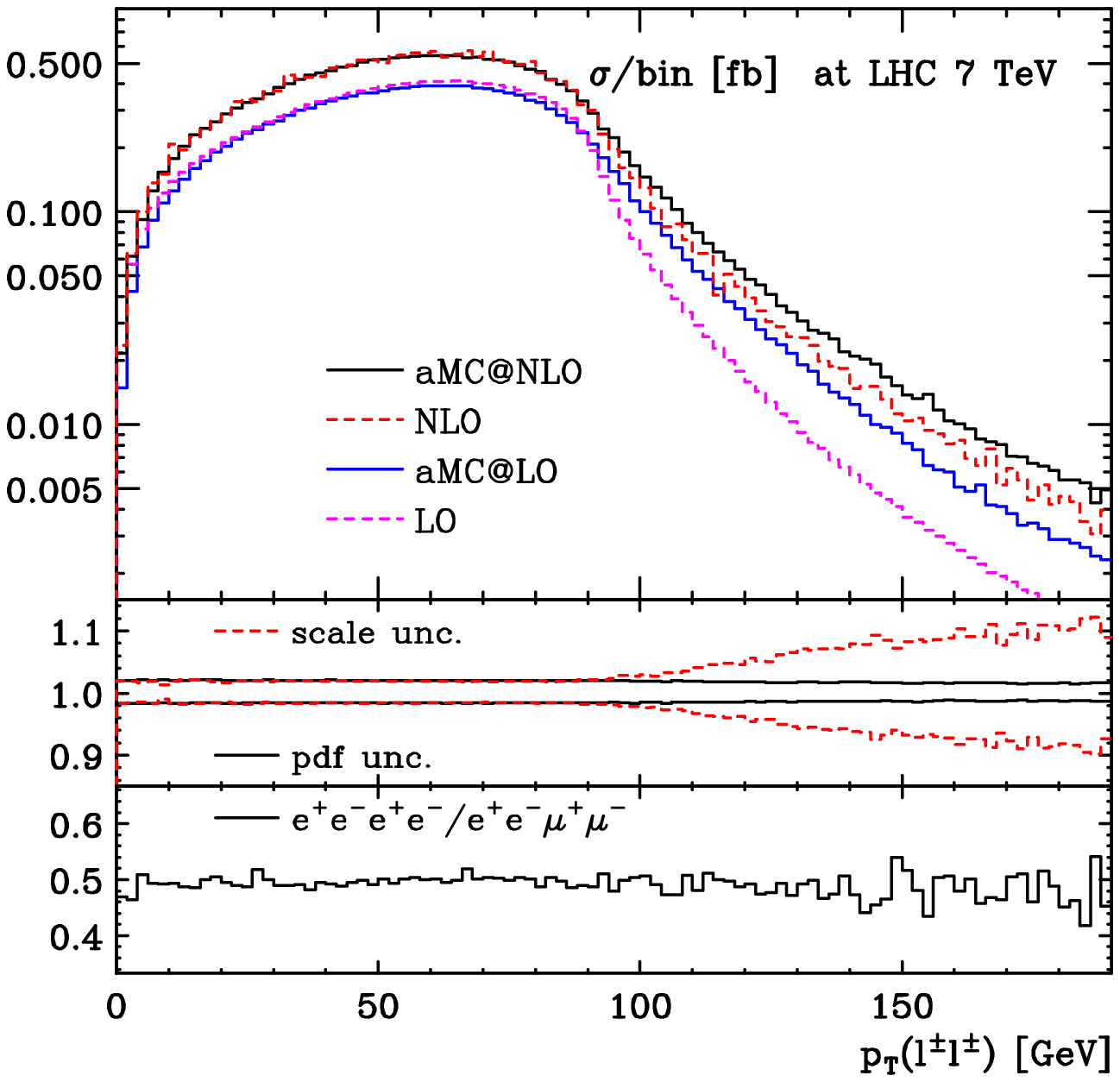}
\caption{Same as in fig.~\ref{fig4:rad}, for the inclusive $\pt$ of
the positively-charged leptons (left panel), and the inclusive $\pt$ 
of the same-charge lepton pairs (right panel), both with $Z$-id cuts.}
\label{fig4:pt}
\end{figure}

\begin{figure}[t]
\centering
\includegraphics[width=0.49\textwidth]{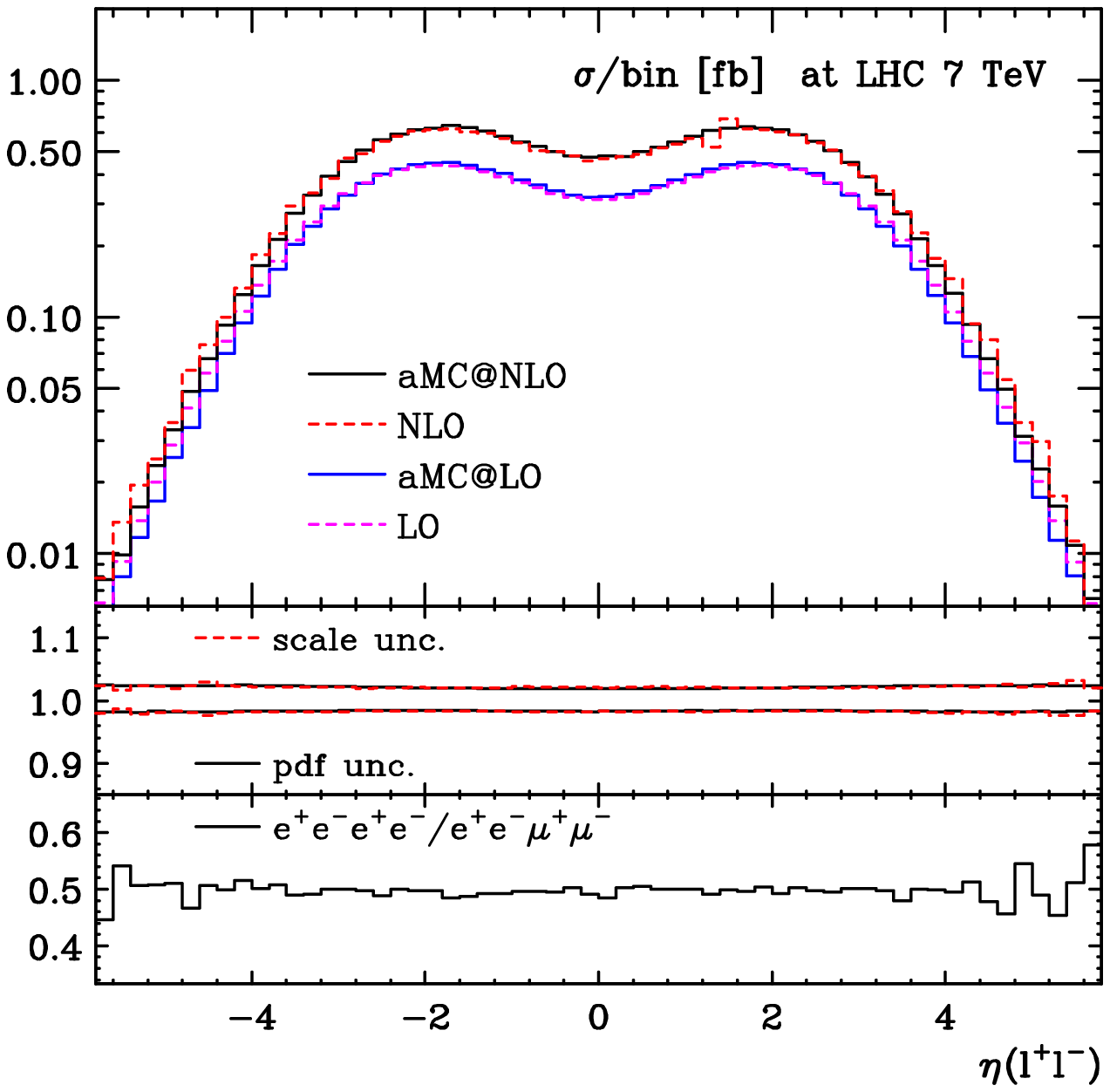}
\includegraphics[width=0.49\textwidth]{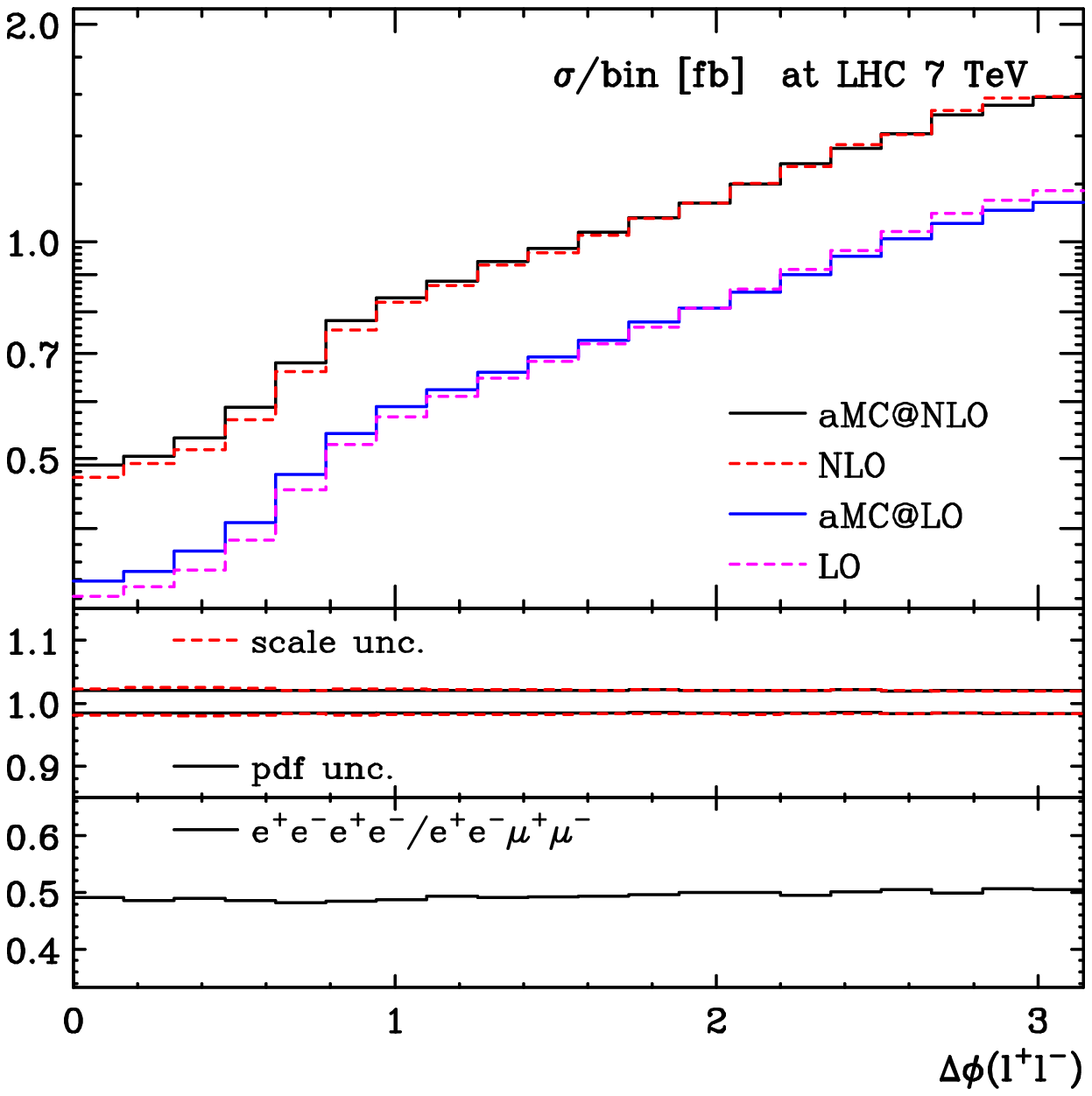}
\caption{As in fig.~\ref{fig4:rad}, for the inclusive $\eta$
of the opposite-charge, $Z$-id matched lepton pairs (left panel),
and the inclusive $\Delta\phi$ distance of the opposite-charge, 
non-$Z$-id matched lepton pairs (right panel).
}
\label{fig4:corr}
\end{figure}

We now turn to discussing the case of differential distributions,
and we start by considering the ${\cal O}(\as^0)$ and ${\cal O}(\as)$
contributions. We shall later address the (generally very small) 
differences between \HW\ and \pythiasix, and here we shall use only \HW\ in 
order to be definite. The aim of figs.~\ref{fig4:rad}--\ref{fig4:corr}
is that of assessing the perturbative behaviour of our predictions.
In the main frame, these figures present the results for the $\epem\mpmm$ 
channel obtained with \amcatnlo\ (solid black), NLO (dashed red), \amcatlo\ 
(solid blue), LO (dashed magenta). The middle insets show the scale 
(dashed red) and PDF (solid black) uncertainties, both with \amcatnlo,
and given as ratios over the central values. Finally, the lower insets 
present the ratios
\beqn
\frac{d\sigma(\epem\epem)}{dO}\Big/\,
\frac{d\sigma(\epem\mpmm)}{dO}\,,
\label{ratofchs}
\eeqn
with $O$ any of the observables considered, as computed by \amcatnlo.

%%%%%%%%%%%%%%%%%%%%%%%%%%%%%%%%%%%%%%%%%%%%%%%%%%%%%%%%%%%%%%%%%%%%%%%%%%%%
%%%%%%%%%%%%%%%%%%%%%%%%%%%%%%%%%%%%%%%%%%%%%%%%%%%%%%%%%%%%%%%%%%%%%%%%%%%%
Figure~\ref{fig4:rad} displays two observables constructed with the
sum of the four-momenta of the four leptons, the invariant mass (left
panel) and the transverse momentum (right panel). These have very different 
behaviours w.r.t.~the extra radiation provided by the parton shower, with 
the former being (almost) completely insensitive to it, and the latter (almost)
maximally sensitive to it. In fact, the predictions for the invariant mass
are basically independent of the shower, with NLO (LO) being equal
to \amcatnlo\ (\amcatlo) over the whole range considered. The NLO
corrections amount largely to an overall rescaling, with a very minimal
tendency to harden the spectrum. The four-lepton $\pt$, on the other
hand, is a well known example of an observable whose distribution at 
the parton-level LO is a delta function (in this case, at $\pt=0$).
Radiation, be it through either showering or hard emission provided
by real matrix elements in the NLO computation, fills the phase space
with radically different characteristics, \amcatlo\ being meaningful
at small $\pt$ and NLO parton level at large $\pt$ -- \amcatnlo\ correctly
interpolates between the two. The different behaviours under extra radiation
of the two observables shown in fig.~\ref{fig4:rad} is reflected in the
scale uncertainty: while in the case of the invariant mass the band
becomes very marginally wider towards large $M(\epem\mpmm)$ values,
the corresponding effect is dramatic in the case of the transverse
momentum. This is easy to understand from the purely perturbative point 
of view, and is due to the fact that, in spite of being ${\cal O}(\as)$
for any $\pt>0$, the transverse momentum in this range is effectively
an LO observable (the NLO effects being confined to $\pt=0$).
The matching with shower blurs this picture, and in particular it
gives rise to the counterintuitive result where the scale dependence
increases, rather than decreasing, when moving towards large 
$\pt$~\cite{Torrielli:2010aw}. Finally, the lower insets of 
fig.~\ref{fig4:rad} display the ratio defined in eq.~(\ref{ratofchs}) 
which, in agreement with the results of table~\ref{tab:xsec}, is equal to
one half in the whole kinematic ranges considered. The only exception is
the small invariant mass region, where off-resonance effects
become relevant.

A gauge-invariant way to suppress off-resonance effects, and to
select doubly-resonant contributions, is that of imposing:
\beqn
\abs{M(\ell^+\ell^-)-m_{\sss Z}}\le 10~{\rm GeV}
\label{Zidcut}
\eeqn
on all equal-flavour lepton pairs; we call the cut of eq.~(\ref{Zidcut})
the $Z$-id cut. Lepton pairs that pass the $Z$-id cut are called
$Z$-id matched, and can be roughly seen as coming from the decay
of a (generally off-shell) $Z$ boson. 
While in the case of the $\epem\mpmm$ channel there
is only one way to choose two same-flavour lepton pairs, there are two 
different pairings in $\epem\epem$ production. In the case both of these
pairings result in lepton pairs that fulfill eq.~(\ref{Zidcut}), we choose 
that with the smallest pair invariant mass, and assign the $Z$-id matched
pairs according to this choice; in practice, this is a rare event. By 
imposing the $Z$-id cuts the $M(\ell^+\ell^-\ell^{(\prime)+}\ell^{(\prime)-})$
distribution falls steeply below threshold and gets no contributions 
below $160$~GeV.

%%%%%%%%%%%%%%%%%%%%%%%%%%%%%%%%%%%%%%%%%%%%%%%%%%%%%%%%%%%%%%%%%%%%%%%%%%%%
%%%%%%%%%%%%%%%%%%%%%%%%%%%%%%%%%%%%%%%%%%%%%%%%%%%%%%%%%%%%%%%%%%%%%%%%%%%%
In fig.~\ref{fig4:pt} we present two transverse momentum distributions,
relevant to the positively-charged leptons (left panel), and to same-charge 
lepton pairs (right panel); hence, there are two entries in each histogram
for any given event. These results are obtained by applying the $Z$-id cuts,
but we have in fact verified that without such cuts we obtain exactly
the same patterns. In the case of the $\pt$ of the individual lepton,
the \amcatnlo\ (\amcatlo) prediction is fairly close to the NLO (LO)
one, but tends to be slightly harder, owing to the extra radiation
generated by the shower. This effect is more pronounced at the LO 
than at the NLO, which is the sign of a behaviour consistent with
perturbation theory expectations. In fact, at the LO all hadronic 
transverse momentum is provided by the shower, while at the NLO this
is not the case; therefore, at the NLO the shower will have less necessity 
to ``correct'' the prediction obtained at the parton level, a tendency
which is naturally embedded in a matching prescription such as \amcatnlo.
The scale dependence is quite small over the whole range in $\pt$, but
tends to grow larger towards larger $\pt$. This effect has the same
origin as that observed in the right panel of fig.~\ref{fig4:rad},
but it is much more moderate than there. This is due to the fact that
in the present case the whole range in $\pt$ is associated with
complete NLO corrections. The PDF uncertainty is seen to be similar to
or slightly smaller than that due to scale variation; parton densities
are well determined in the $x$ range probed here. Finally, there is
no difference between the two leptonic channels for this observable;
as already mentioned above, this conclusion is independent of whether
one applies the $Z$-id cuts. The $\pt$ of the lepton pairs shown in
the right panel of fig.~\ref{fig4:pt} follows the same pattern as
the one we have just discussed, but the differences between the various
predictions are larger in this case. In particular, \amcatlo\ is closer
to NLO than to LO, which is a consequence of the more important role
played by extra radiation in this case (as one expects, the present
one being a correlation between two particles rather than a single-inclusive
observable). Again, the closeness of NLO and \amcatnlo\ results shows
the desired perturbative behaviour. The more significant impact of
extra radiation on this variable is reflected in the slightly larger
scale dependence at large $\pt$'s w.r.t.~what happens for the transverse
momentum of the individual leptons discussed before. The two leptonic
channels agree well, also when removing the $Z$-id cuts.

%%%%%%%%%%%%%%%%%%%%%%%%%%%%%%%%%%%%%%%%%%%%%%%%%%%%%%%%%%%%%%%%%%%%%%%%%%%%
%%%%%%%%%%%%%%%%%%%%%%%%%%%%%%%%%%%%%%%%%%%%%%%%%%%%%%%%%%%%%%%%%%%%%%%%%%%%
Figure~\ref{fig4:corr} shows two observables constructed after applying the
$Z$-id cuts, namely the pseudorapidity of lepton pairs with opposite charge
which are also $Z$-id matched (left panel; this is then the pseudorapidity of
would-be $Z$ bosons), and the azimuthal distance between leptons of opposite
charge which are not $Z$-id matched (right panel; thus, these are leptons
emerging from different would-be $Z$ bosons). As in the case of
fig.~\ref{fig4:pt}, there are two entries in each histogram for any given
event. These two observables are dominated by small transverse momenta, and
therefore it is not suprising that, at both ${\cal O}(\as^0)$ and ${\cal
O}(\as)$, the predictions are quite independent of whether a shower is
generated or not. Slight differences can be seen in the case of the
$\Delta\phi$ distribution, which is indeed known to be more sensitive than
pseudorapidity to extra radiation.  The small-$\pt$ dominance ensures that
scale and PDF uncertainties are flat over the whole kinematic ranges, and of
the order of those relevant to total cross section.

We now discuss the impact of the ${\cal O}(\as^2)$ $gg$ channel
on our predictions. The argument for considering such a channel,
despite its being of the same perturbative order as all other
NNLO contributions which cannot be included, is the dominance
of its parton luminosity over those of the $q\qb$ and $qg$ channels.
This dominance grows stronger with decreasing final-state invariant 
masses, and hence the ${\cal O}(\as^2)$ versus NLO comparison is
significantly influenced by the cut in eq.~(\ref{Mcut}) -- by
lowering such a cut, the relative importance of the $gg$ contribution
will grow bigger than the 5\%-ish reported in table~\ref{tab:xsec}.
We also discuss in the following the differences that arise when
matching our calculation to \pythiasix\ rather than to \HW. We remind the
reader that, depending on input parameters, \PY\ is rather effective
in producing radiation in the whole kinematically-accessible phase space.
This is not particularly useful in the context of a matched computation,
where hard radiation is provided (in a way fully consistent with perturbation
theory) by the underlying real-emission matrix elements. Therefore, we have 
set the maximum virtuality in \PY\ equal to the four-lepton invariant mass.
For consistency, this setting has been used also when showering
the $gg$-initiated contribution.

%%%%%%%%%%%%%%%%%%%%%%%%%%%%%%%%%%%%%%%%%%%%%%%%%%%%%%%%%%%%%%%%%%%%%%%%%%%%
\begin{figure}[t]
\centering
\includegraphics[width=0.49\textwidth]{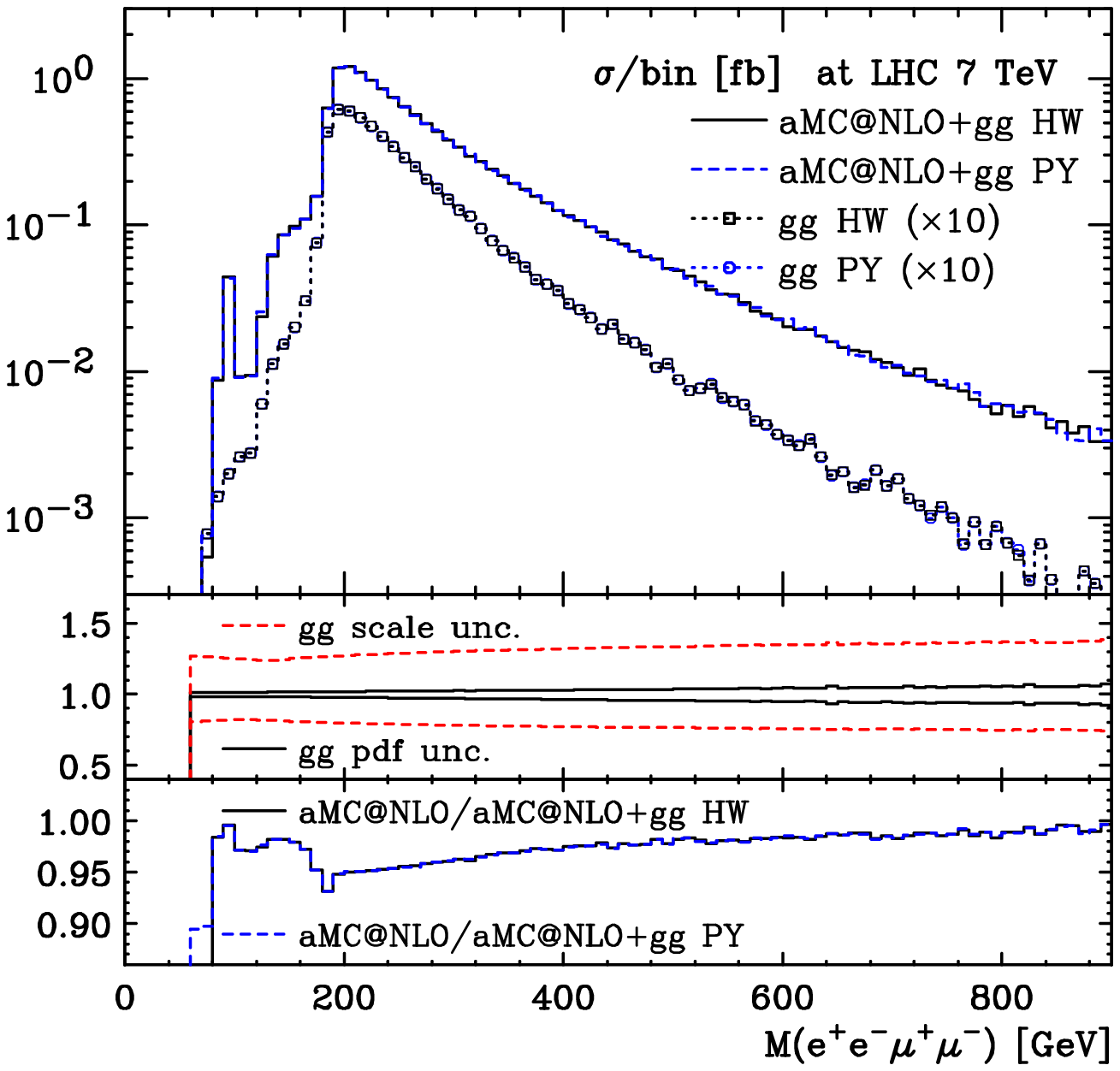}
\includegraphics[width=0.49\textwidth]{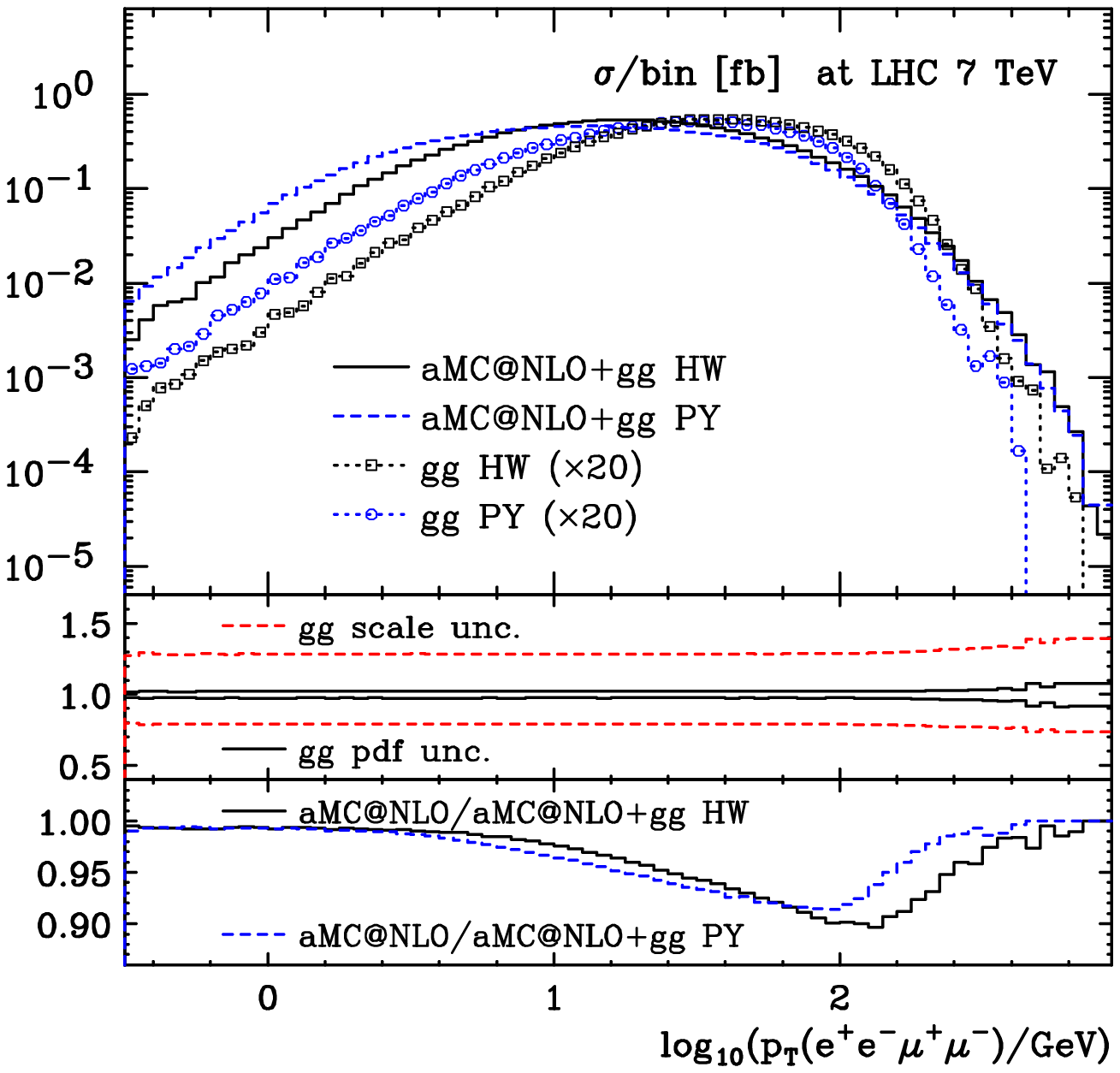}
\caption{Same observables as in fig.~\ref{fig4:rad}, for \amcatnlo+$gg$
\HW\ (solid black) and \PY\ (dashed blue) results. The rescaled $gg$
contributions with \HW\ (open black boxes) and \PY\ (open blue circles)
are shown separately. Middle insets: scale (dashed red) and PDF (solid black)
fractional uncertainties. Lower insets: \amcatlo/(\amcatnlo+$gg$) with 
\HW\ (solid black) and \PY\ (dashed blue).}
\label{fig5:rad}
\end{figure}
\begin{figure}[t]
\centering
\includegraphics[width=0.49\textwidth]{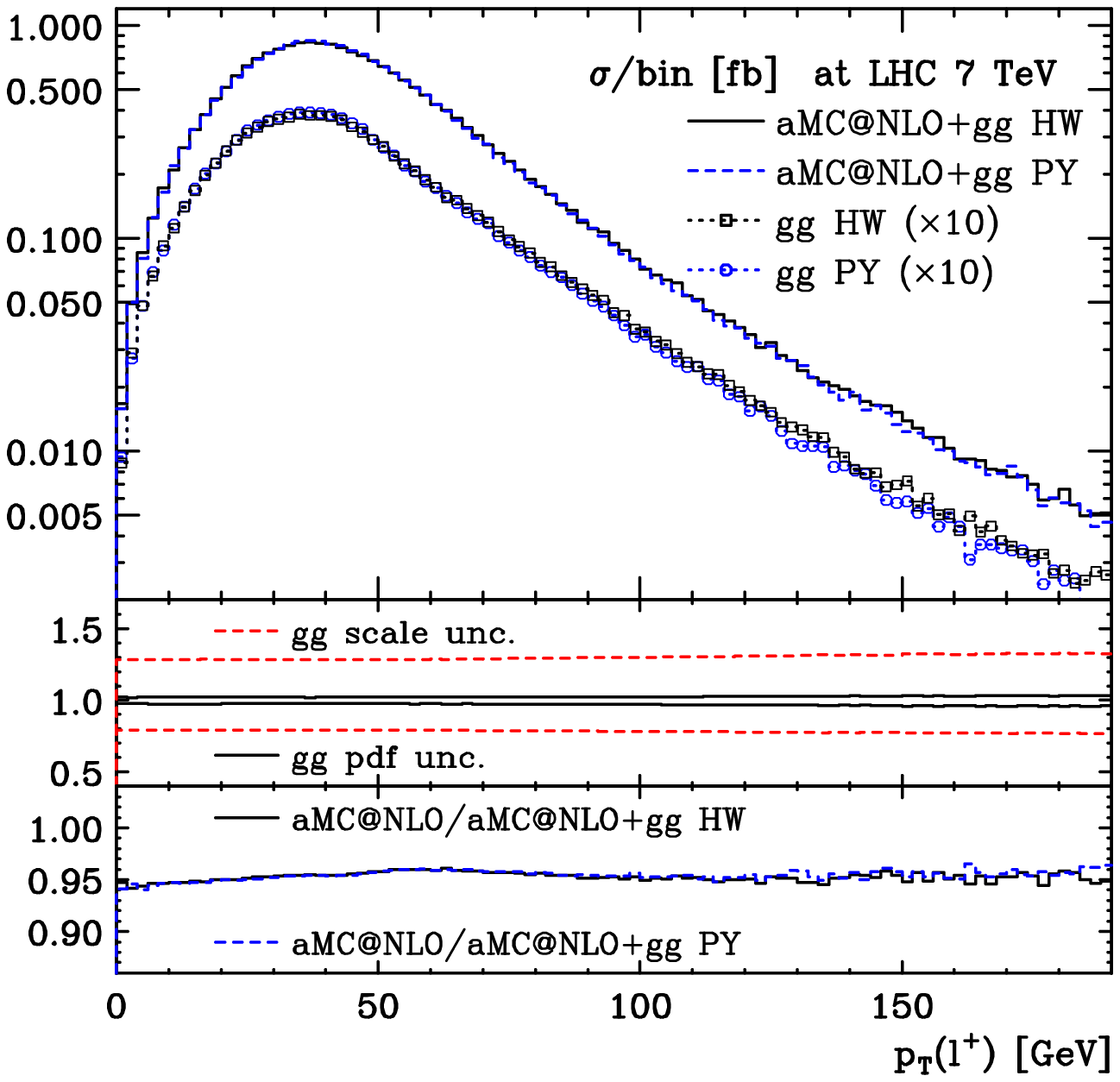}
\includegraphics[width=0.49\textwidth]{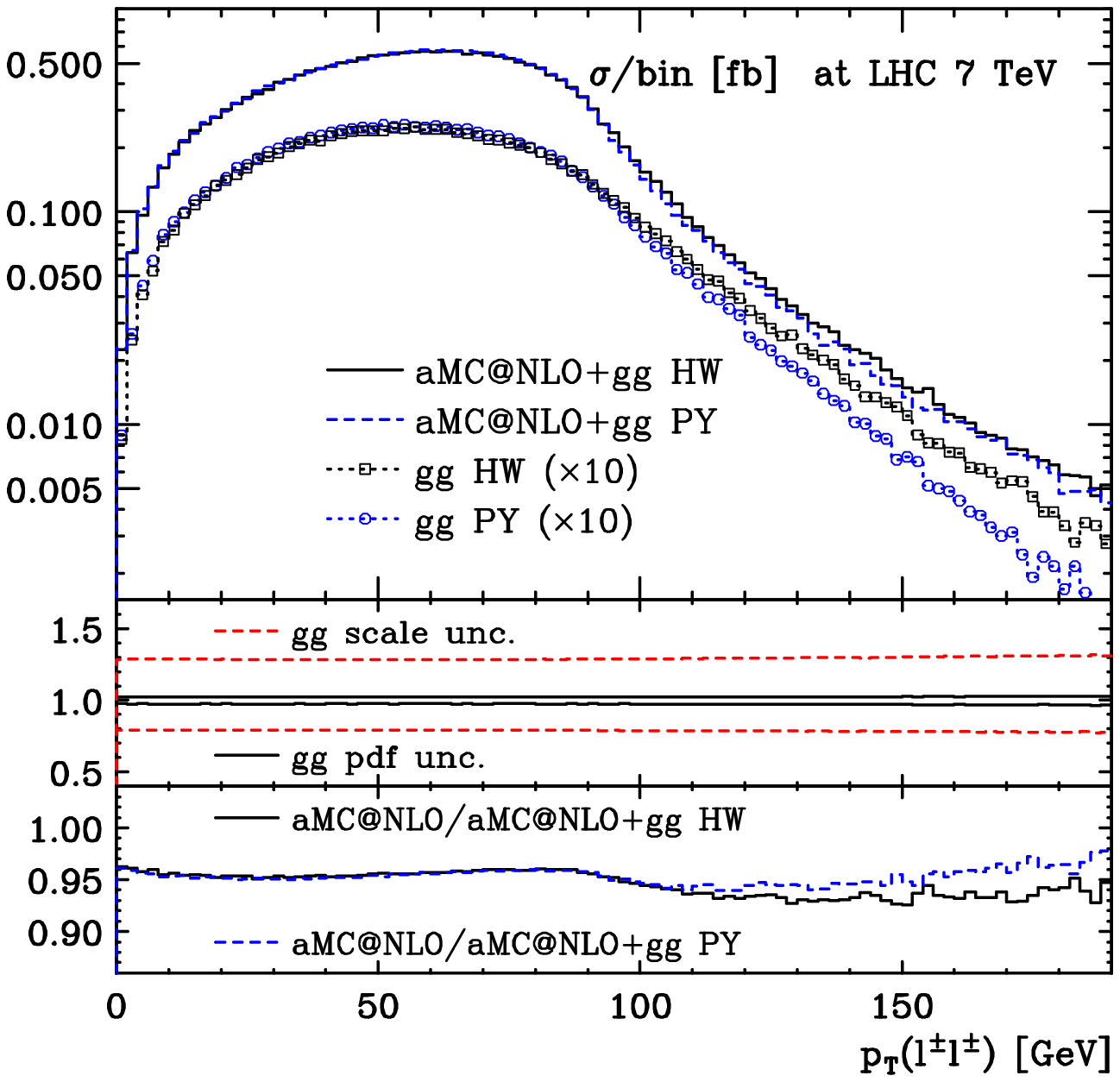}
\caption{Same as in fig.~\ref{fig5:rad}, for the observables introduced
in fig.~\ref{fig4:pt}.}
\label{fig5:pt}
\end{figure}
\begin{figure}[t]
\centering
\includegraphics[width=0.49\textwidth]{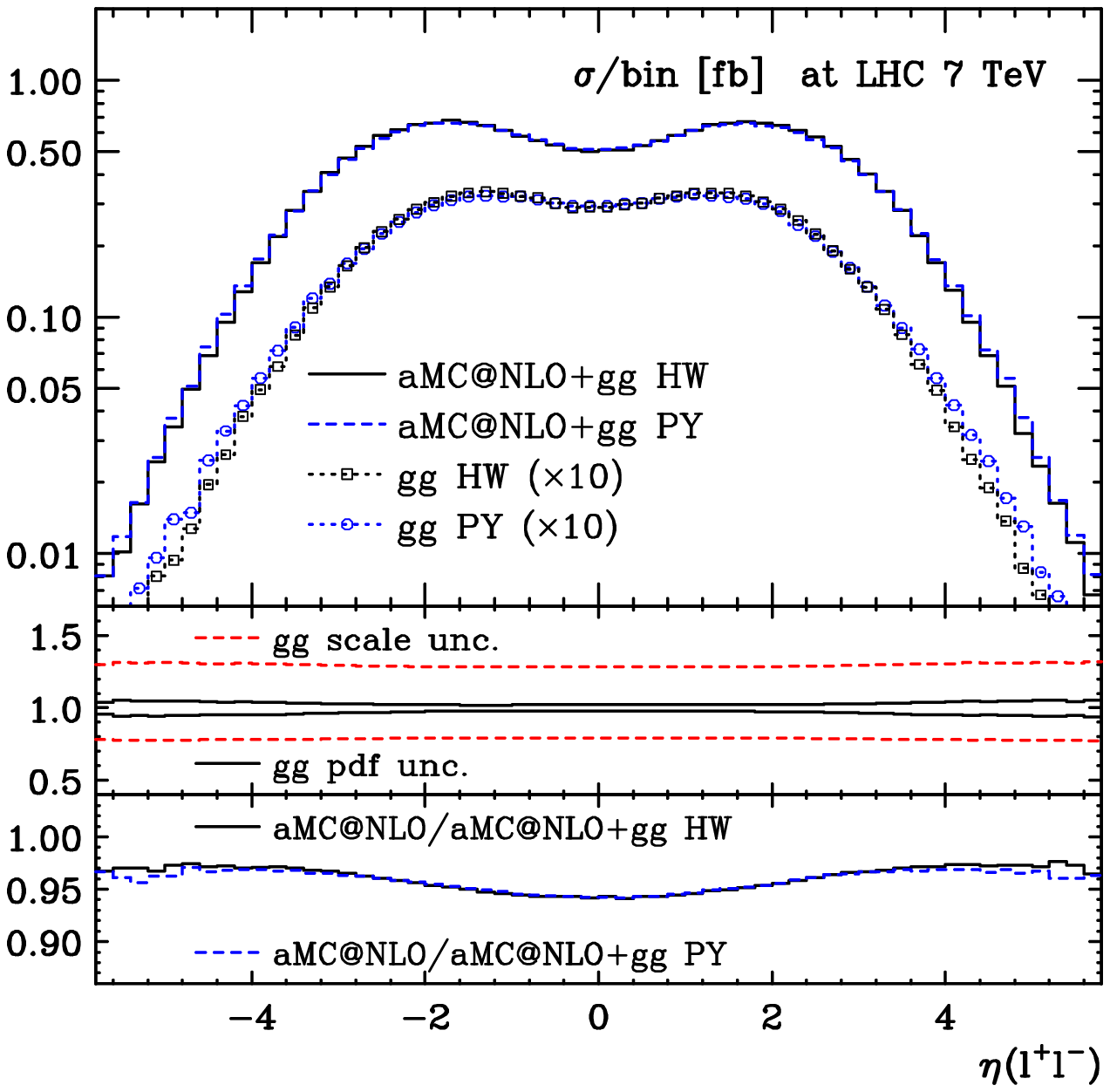}
\includegraphics[width=0.49\textwidth]{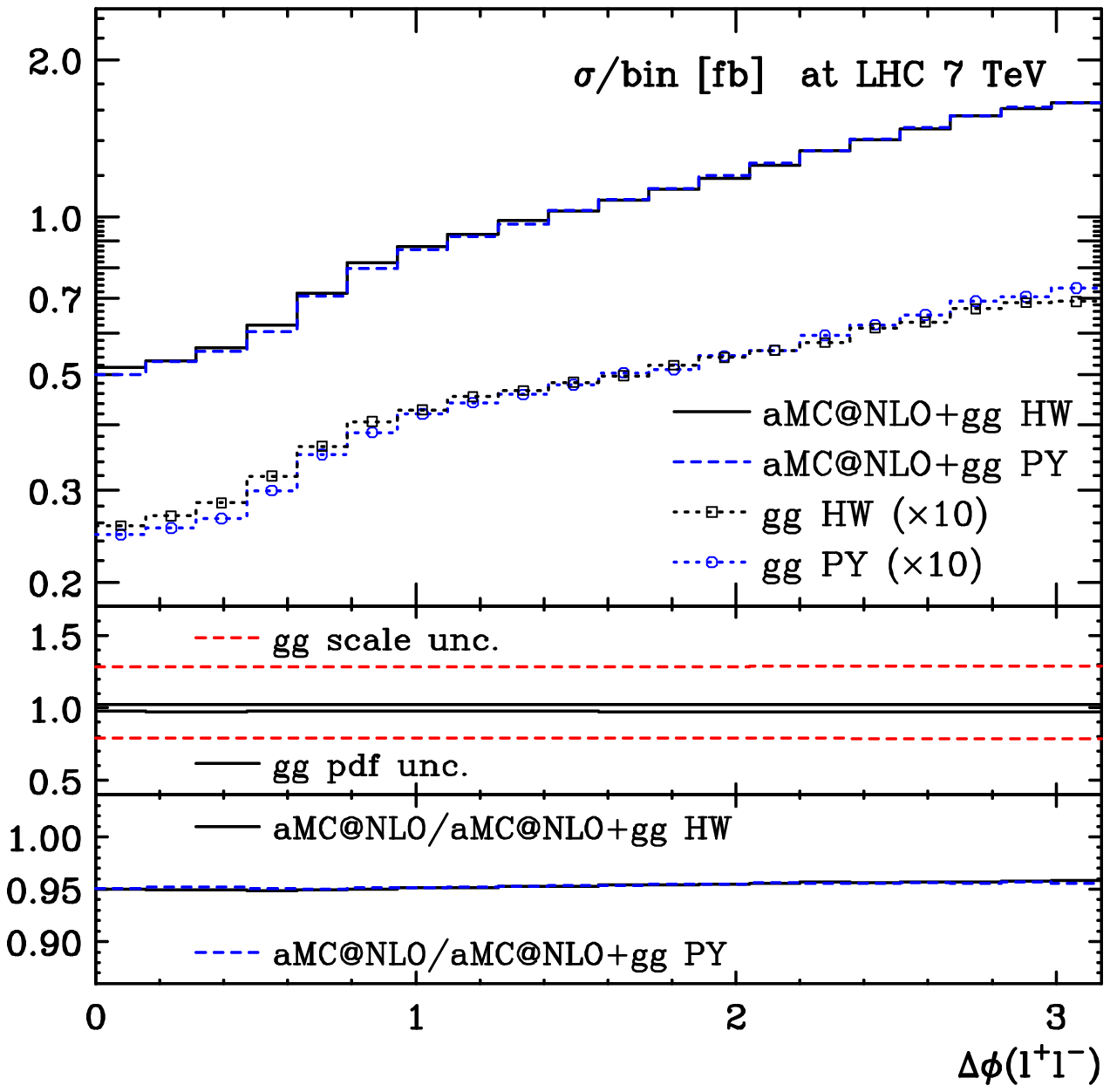}
\caption{Same as in fig.~\ref{fig5:rad}, for the observables introduced
in fig.~\ref{fig4:corr}.}
\label{fig5:corr}
\end{figure}

%%%%%%%%%%%%%%%%%%%%%%%%%%%%%%%%%%%%%%%%%%%%%%%%%%%%%%%%%%%%%%%%%%%%%%%%%%%%
Figures~\ref{fig5:rad}, \ref{fig5:pt} and \ref{fig5:corr} present
the same observables as figs.~\ref{fig4:rad}, \ref{fig4:pt}
and~\ref{fig4:corr} respectively. In the main frame, 
we show the \amcatnlo\ predictions
plus the $gg$ contribution (including shower), as resulting from
\HW\ (solid black) and \PY\ (dashed blue) -- we shall call these
predictions \amcatnlo+$gg$ for brevity. The dashed-plus-symbol
histograms are the $gg$ contributions (open black boxes: \HW;
open blue circles: \PY), rescaled in order to fit on the 
same scale as \amcatnlo+$gg$; the rescaling factors are
equal either to 10 or 20, as indicated in the figures -- they are 
chosen so as to render the comparisons of shapes as transparent as 
possible. The middle insets show the fractional scale (dashed red) 
and PDF (black solid) uncertainties of the $gg$ contribution
alone (i.e., the extrema divided by the central $gg$ prediction);
those shown are computed with \HW, but \PY\ gives identical results.
The lower insets present the ratios \amcatnlo$/$(\amcatnlo+$gg$) for 
\HW\ (solid black) and \PY\ (dashed blue).

%%%%%%%%%%%%%%%%%%%%%%%%%%%%%%%%%%%%%%%%%%%%%%%%%%%%%%%%%%%%%%%%%%%%%%%%%%%%
%%%%%%%%%%%%%%%%%%%%%%%%%%%%%%%%%%%%%%%%%%%%%%%%%%%%%%%%%%%%%%%%%%%%%%%%%%%%
The \HW\ and \PY\ results for the four-lepton invariant mass shown
in the left panel of fig.~\ref{fig5:rad} are identical in the case 
of the $gg$ contribution, and practically so in the case of \amcatnlo.
This is because both event generators fix the invariant mass of the
final-state system when doing initial-state showers; hence, any
difference in the four-lepton invariant mass can only arise from
the real-emission contribution to \amcatnlo, where the final-state
system does not coincide with the four leptons. The $gg$ channel
has a softer distribution than the $q\qb+qg$ ones, since by moving
towards large masses one probes larger values of the Bjorken $x$'s,
where quark parton luminosities are relatively more important than
at threshold. In the small-invariant-mass region, the $gg$ contribution
does not exhibit any peak structure, owing to the lack of singly-resonant
diagrams. Ultimately, the only non-negligible effects are around the
peak region, where they are larger than 5\%. The four-lepton transverse
momentum is marginally softer with \pythiasix\ than with \HW, an effect that
is more pronounced in the $gg$ contribution (because of a larger
amount of radiation there w.r.t.~the case of the $q\qb$ and $qg$ 
channels, radiation being on average softer in \pythiasix\ than in \HW).
On the other hand, this larger amount of radiation implies that 
the $gg$ channel is significantly harder than \amcatnlo, and is
about 10\% of the \amcatnlo+$gg$ result at $\pt={\cal O}(100~{\rm GeV})$.
The scale uncertainty shows a behaviour typical of an LO computation,
and is vastly different from that of fig.~\ref{fig4:rad}; it is 
constant to a good approximation, which is due to a decreasing $\muR$
dependence with increasing $\pt$, compensated by an increasing $\muF$
dependence.

As was already pointed out in the case of fig.~\ref{fig4:pt}, the
observables presented in fig.~\ref{fig5:pt} are quite insensitive
to extra radiation, and especially so for the single-inclusive
lepton $\pt$. Therefore, the similarity between the predictions
obtained with \HW\ and \PY\ has to be expected. To a very good
extent, the $gg$ contribution has the same shape as the \amcatnlo\ one.
Things are marginally different in the case of the pair transverse
momentum presented in the right panel; however, any difference
between \HW\ and \PY\ is not significant, being much smaller than
the scale uncertainty of the \amcatnlo\ contribution alone
(shown in fig.~\ref{fig4:pt}). Likewise, the shape change induced
in the \amcatnlo\ prediction by the $gg$ contribution is very minor.

%%%%%%%%%%%%%%%%%%%%%%%%%%%%%%%%%%%%%%%%%%%%%%%%%%%%%%%%%%%%%%%%%%%%%%%%%%%%
%%%%%%%%%%%%%%%%%%%%%%%%%%%%%%%%%%%%%%%%%%%%%%%%%%%%%%%%%%%%%%%%%%%%%%%%%%%%
The small-$\pt$ dominance for the observables displayed in
fig.~\ref{fig5:corr} implies that again the \HW\ and \PY\ results
are very similar. As far as the $gg$ contribution is concerned,
its shape is almost identical to that of \amcatnlo\ in the case
of $\Delta\phi$. On the other hand, the lepton-pair pseudorapidity
turns out to be significantly more central than that of the 
$q\qb+qg$ channels; the impact on the overall \amcatnlo+$gg$ 
shape is however modest for the invariant-mass cuts considered here.

\section{Conclusions\label{sec:concl}}

In this paper, we have used the \amcatnlo\ framework to study the 
hadroproduction of four leptons, by considering, in particular, the process 
mediated by the exchange of two $Z$ or $\gamma$ vector bosons. Di-bosons 
final states are relevant in both Higgs searches and new-physics studies, 
where precise predictions are an important ingredient to cope with the 
overwhelming irreducible Standard Model backgrounds. We have included 
in our calculation all contributions which are expected to be important
for accurate phenomenological analyses, such as singly-resonant diagrams, 
exact spin correlations and off-shell effects, $Z/\gamma^*$ interference,
and $gg$-initiated subprocesses.  Our computation being fully
automated, any four-lepton final states can be studied in the very same way,
including those that feature $W/Z$ interference effects such as
$\epem\nu_e\bar{\nu}_e$.

In any realistic study, precise calculations are as important as the
determination of the associated theoretical uncertainties. For this reason,
we have also given in this work practical applications of the fully automatic 
procedures implemented to this aim in \amcatnlo, namely:
\begin{itemize}
\item[{\em i)}] The matching of the hard process to both \pythiasix\ and
\HW, paving the way to systematic studies of the influence of different 
showers on any observable with \mcatnlo\footnote{The matching with
\herwigpp\ and {\sc Pythia8} will become available as well.};
\item[{\em ii)}] A framework for studying both scale and PDF 
uncertainties by a simple (and CPU-wise costless) reweighting procedure 
of the unweighted-event samples.
\end{itemize}
As for item {\em i)}, this paper provides the first working example of
matching with \pythiasix\ for a kinematically non-trivial process.  
As for item {\em ii)}, we stress that our procedure allows one to obtain 
precise results with very limited statistics. In fact, the central results 
and all its variations are correlated, since the same events (up to the 
weights) are used. This is manifest in the smoothness of the scale and PDF 
fractional uncertainty plots presented in this paper, which may be compared 
with, for example, those presented in ref.~\cite{Torrielli:2010aw}, which 
had been obtained by rerunning the code.

As far as our phenomenological results are concerned, they can be summarized 
as follows. The impact of NLO corrections on total rates is +40$\%$;
for differential distributions, they generally correspond to an overall
rescaling, but there are cases where they give non-trivial kinematics
effects. The uncertainties due
to PDFs are of the order of 2$\%$. The scale dependence we obtain by varying
$\muR$ and $\muF$ independently is also of the order of 2\% for the 
$q\qb/qg$ channels, and of the order of 20\% in the case of the $gg$-initiated
subprocess. Parton-level NLO and fully showered NLO distributions are 
generally in good agreement, except in the very few cases in which the
former is not expected to give sensible predictions.
These findings are rather independent of possible
$Z$-identification cuts applied to enhance doubly-resonant contributions.
Generally speaking, tiny differences are observed between \HW\ and \PY, 
with the exception of distributions dominated by resummation effects,
where they are more pronounced. The contribution of the $gg$ channel is 
of the order of $5\%$ of the total with a cut of $30$ GeV applied to 
unlike-sign lepton pairs.

We have explicitly verified that the inclusion of the $gg\to H\to ZZ$
contribution (not considered for producing the results presented here)
is straightforward, and does not increase significantly the running
time. This opens the way to simulating Higgs production with four-lepton
decay channels by including the signal/background interference.
Clearly, for this to be phenomenologically sensible, NLO corrections 
to the $gg\to H$ signal should to be included as well, which could be 
achieved in several ways, with different levels  of approximation and 
automation~\cite{Frixione:2008yi,Dittmaier:2011ti,Campbell:2011cu}. 

As a last remark, we point out that ready-to-shower four-lepton NLO 
(i.e., resulting from $q\qb/qg$-initiated processes) event 
files are available at the \amcatnlo\ web page {\tt http://amcatnlo.cern.ch},
for all possible (massless) lepton flavour combinations\footnote{Processes
that feature two $W$ propagators interfere at the NLO with $tW$ production.
We have excluded the latter contribution by setting the $b$-quark PDF 
equal to zero.}.

\section{Acknowledgments}
This research has been supported by the Swiss National Science Foundation
(SNF) under contracts 200020-138206 and 200020-129513, 
by the IAP Program, BELSPO P6/11-P, the
IISN convention 4.4511.10, by the Spanish Ministry of education under contract
PR2010-0285, and in part by the US National Science Foundation under Grant
No. NSF PHY05-51164. F.M. and R.P. thank the financial support of the MEC
project FPA2008-02984 (FALCON). R.F. and R.P would like to thank the KITP at
UCSB for the kind hospitality offered during the completion of this paper.

\appendix
\section{Definition of $W$ coefficients for NLO cross sections
\label{sec:wgtNLO}}
We collect in this section the expressions of the coefficients that 
enter eq.~(\ref{Wadef}). These are taken from ref.~\cite{Frederix:2009yq},
whose equation~(n) will be denoted by \MadFKSeq{n} here; the reader 
can find fuller details in that paper. The master equation for the
determination of all the coefficients $\Wa$ is \MadFKSeq{C.14}. 

~\\
\noindent 
$\bullet$ {\em Event}\\
Only the $(n+1)$-body cross section contributes 
to the event. Therefore, from \MadFKSeq{4.29} and \MadFKSeq{6.7} one gets:
\beqn
\hWEz&=&w_{real}\left(\confnpoE\right)\,,
\label{wEdef}
\\
\hWEF&=&0\,,
\\
\hWER&=&0\,,
\eeqn
where 
\beqn
w_{real}(\confnpo)&=&
\frac{\symmnpoij}{\xii(1-\yij)}\,
\frac{\Sigma_{ij}(\confnpo)}{\gs^{2b+2}(\muR)}\,J_{Bj}\,,
\label{wrealdef}
\\
\Sigma_{ij}&=&\left((1-\yij)\xii^2\ampsqnpot\right)
\Sfunij\frac{\JetsB}{\avg}\tPhspnpo\,,
\label{Sigdef}
\eeqn
with $\tPhspnpo=\Phspnpo/\xii$.
The quantity defined in eq.~(\ref{Sigdef}) is the combination of the 
real-emission matrix element, prefactors, and the phase-space factor 
that is routinely used in the FKS subtraction. We point out that
according to our conventions the matrix elements include the coupling
constant, hence eq.~(\ref{wrealdef}) is independent of $\gs$, since
we understand that the factorization scale $\muR$ there is computed
according to the kinematic configuration $\confnpo$. The coupling-constant 
dependence has been factored out in eq.~(\ref{Wadef}).

~\\
\noindent 
$\bullet$ {\em Collinear counterevent}\\
The coefficient $\WC$ receives contributions from the $(n+1)$-body and
the degenerate $(n+1)$-body cross sections. The latter can be read
from \MadFKSeq{4.41} (to be definite, we assume that the FKS sister
is the parton coming from the left). One obtains:
\beqn
\hWCz&=&-w_{real}\left(\confnpoC\right)+
\frac{1}{\gs^{2b+2}(\muRC)}
\frac{\as(\muRC)}{2\pi}
\Bigg\{\APdamp_{\ident_{1\oplus\bar{i}}\ident_1}^{(0)}(1-\xii)
\left[\frac{1}{\xii}\log\frac{s\deltaI}{2\QESCt}+
2\frac{\log\xii}{\xii}\right]
\nonumber \\*&&
\phantom{d\bar{\sigma}_{i1}^{(n+1)}(\proc;k_1,k_2)=\asotwopi}
-\APdamp_{\ident_{1\oplus\bar{i}}\ident_1}^{(1)}(1-\xii)\frac{1}{\xii}
-K_{\ident_{1\oplus\bar{i}}\ident_1}(1-\xii)\Bigg\}
\nonumber \\*&&\phantom{-w_{real}\left(\confnpoC\right)}\times
\ampsqnt\left(\confnpoC\right)
\frac{\JetsB}{\avg}{\cal F}_{\Phi}\tPhspnpo\left(\confnpoC\right)J_{Bj}\,,
\label{wdegCdef}
%%%%\eeqn
%%%%\beqn
\\
\hWCF&=&-\frac{1}{\gs^{2b+2}(\muRC)}
\frac{\as(\muRC)}{2\pi}
\APdamp_{\ident_{1\oplus\bar{i}}\ident_1}^{(0)}(1-\xii)
\frac{1}{\xii}
\nonumber \\*&&\phantom{aaaaaa}\times
\ampsqnt\left(\confnpoC\right)
\frac{\JetsB}{\avg}{\cal F}_{\Phi}\tPhspnpo\left(\confnpoC\right)J_{Bj}\,,
\label{wdegFCdef}
\\
\hWCR&=&0\,.
\eeqn
The normalization factor ${\cal F}_{\Phi}$ in eqs.~(\ref{wdegCdef}) 
and~(\ref{wdegFCdef}) is due to the fact that the degenerate $(n+1)$-body 
contribution is integrated together with the pure $(n+1)$-body one, but 
is originally defined with a quasi-$n$-body measure (see \MadFKSeq{4.41}).

~\\
\noindent 
$\bullet$ {\em Soft counterevent, $n$-body contribution, and Born}\\
The coefficient $\WS$ receives contributions from the $(n+1)$-body and
the $n$-body cross sections. The latter is the sum of the Born
(\MadFKSeq{4.4}), collinear (\MadFKSeq{4.5}), soft (\MadFKSeq{4.12})
and finite virtual (\MadFKSeq{4.14}) contributions, plus the last term
on the r.h.s.~of \MadFKSeq{C.14}, which we shall call the RGE term here.
The expression for the finite part of the virtual contribution that includes 
the full scale dependence must be read from \MadFKSeq{C.13}. 
In ref.~\cite{Frederix:2009yq} we have used the Ellis-Sexton scale 
wherever possible, which now renders it easy to extract the factorization
and renormalization scale dependences -- more specifically, these dependences
are present only in one term in \MadFKSeq{4.5}, in one term in 
\MadFKSeq{C.13}, and in the RGE term. This implies:
\beqn
\hWSz&=&-w_{real}\left(\confnpoS\right)+
\frac{1}{\gs^{2b+2}(\muRS)}
\frac{\as(\muRS)}{2\pi}\Bigg[
{\cal Q}\left(\mu=\QESS\right)
\ampsqnt\left(\confnpoS\right)
\nonumber\\*&&\phantom{aaaa}
+\frac{\as^b(\muRS)}{\as^b(\QESS)}
\vampsqnlF\left(\QESSt,\QESSt,\QESSt;\confnpoS\right)
\nonumber \\*&&\phantom{aaaa}
+\sum_{k,l}\eikint_{kl}^{(m_k,m_l)}\ampsqnt_{kl}\left(\confnpoS\right)
\Bigg]
\frac{\JetsB}{\avg}
{\cal F}_{\Phi}^{(B)}\tPhspnpo\left(\confnpoS\right)J_{Bj}\,,
\label{wSndef}
\\
\hWSF&=&-\frac{1}{\gs^{2b+2}(\muRS)}
\frac{\as(\muRS)}{2\pi}
\ampsqnt\left(\confnpoS\right)
\frac{\JetsB}{\avg}
{\cal F}_{\Phi}^{(B)}\tPhspnpo\left(\confnpoS\right)J_{Bj}
\nonumber \\*&&\phantom{aa}\times
\left(\gamma(\ident_1)+2C(\ident_1)\log\xicut
+\gamma(\ident_2)+2C(\ident_2)\log\xicut\right)\,,
\label{wSFndef}
\\
\hWSR&=&\frac{1}{\gs^{2b+2}(\muRS)}
\frac{\as(\muRS)}{2\pi}
\;2\pi\beta_0\,b
\nonumber \\*&&\phantom{aa}\times
\ampsqnt\left(\confnpoS\right)
\frac{\JetsB}{\avg}
{\cal F}_{\Phi}^{(B)}\tPhspnpo\left(\confnpoS\right)J_{Bj}\,,
\label{wSRndef}
\\
\hWB&=&\frac{\ampsqnt\left(\confnpoS\right)}{\gs^{2b}(\muR)}
\frac{\JetsB}{\avg}
{\cal F}_{\Phi}^{(B)}\tPhspnpo\left(\confnpoS\right)J_{Bj}\,.
\eeqn
The normalization factor ${\cal F}_{\Phi}^{(B)}$ is the analogue of
that introduced in eq.~(\ref{wdegCdef}), and its form can be found
e.g.~in \MadFKSeq{6.14} (note that it includes an $\Sfunij$ factor). 
In eq.~(\ref{wSndef}), by setting
$\mu=\QESS$ one eliminates the first term on the r.h.s.~of
\MadFKSeq{4.6}; this is included here in eq.~(\ref{wSFndef}).

\noindent 
$\bullet$ {\em Soft-collinear counterevent}\\
Finally, the results for $\WSC$ can be obtained by computing the soft
limit of $\WC$, and by taking into account the fact that the original
equations contained plus distributions. One obtains:
\beqn
\hWSCz&=&w_{real}\left(\confnpoSC\right)
-\frac{1}{\gs^{2b+2}(\muRSC)}
\frac{\as(\muRSC)}{2\pi}
\Bigg\{\APdamp_{\ident_{1\oplus\bar{i}}\ident_1}^{(0)}(1)
\left[\frac{1}{\xii}\log\frac{s\deltaI}{2\QESSCt}+
2\frac{\log\xii}{\xii}\right]
\nonumber \\*&&
\phantom{d\bar{\sigma}_{i1}^{(n+1)}(\proc;k_1,k_2)=\asotwopi}
-\APdamp_{\ident_{1\oplus\bar{i}}\ident_1}^{(1)}(1)\frac{1}{\xii}
-K_{\ident_{1\oplus\bar{i}}\ident_1}(1)\Bigg\}
\nonumber \\*&&\phantom{w_{real}\left(\confnpoSC\right)}\times
\ampsqnt\left(\confnpoSC\right)
\frac{\JetsB}{\avg}{\cal F}_{\Phi}\tPhspnpo\left(\confnpoSC\right)J_{Bj}\,,
\label{wdegSCdef}
\\
\hWSCF&=&\frac{1}{\gs^{2b+2}(\muRSC)}
\frac{\as(\muRSC)}{2\pi}
\APdamp_{\ident_{1\oplus\bar{i}}\ident_1}^{(0)}(1)
\frac{1}{\xii}
\nonumber \\*&&\phantom{aaaaaa}\times
\ampsqnt\left(\confnpoSC\right)
\frac{\JetsB}{\avg}{\cal F}_{\Phi}\tPhspnpo\left(\confnpoSC\right)J_{Bj}\,,
\label{wdegFSCdef}
\\
\hWSCR&=&0\,.
\eeqn

%%May use eq.(2.39) or (2.40) of the single top paper if we want to
%%define the MC weights introduced in eq.~(\ref{xsecMC}).

%%%%%%%%%%%%%%%%%%%%%%%%%%%%%
\bibliography{fourlep}
%%%%%%%%%%%%%%%%%%%%%%%%%%%%%

\end{document}